\documentclass[twocolumn]{aastex6}
\usepackage{amsmath}
\usepackage{url}

\newcommand{\spitzer}{\mbox {\em Spitzer}}
\newcommand{\chandra}{\mbox {\em Chandra}}
\newcommand{\msx}{{\it MSX}}
\newcommand{\um}{$\mu$m}
\newcommand{\hii}{\mbox{\ion{H}{2}~}}
\renewcommand{\deg}{$^\circ$}
\newcommand{\degree}{^{\circ}}
\newcommand{\Msun}{$M_{\sun}$}
\newcommand{\Lsun}{$L_{\sun}$}
\newcommand{\kms}{km~s$^{-1}$}
\newcommand{\peryr}{yr$^{-1}$}
\newcommand{\avsed}{$A_V^{\rm SED}$}

\slugcomment{Accepted for publication in {\em The Astrophysical Journal} --- 15 April 2016}

\shorttitle{Circumstellar Disk Evolution and Accelerating Star Formation in M17 SWex}
\shortauthors{Povich et al.}

\begin{document}

\title{ Rapid Circumstellar Disk Evolution and an Accelerating Star Formation Rate \\ in the Infrared Dark Cloud M17 SWex
}

\author{Matthew S. Povich,\altaffilmark{1}
  Leisa K. Townsley,\altaffilmark{2}
  Thomas P. Robitaille,\altaffilmark{3,4}
  Patrick S. Broos,\altaffilmark{2}
  Wesley T. Orbin,\altaffilmark{2} \\
  Robert R. King,\altaffilmark{5}
  Tim Naylor,\altaffilmark{5}
  \& Barbara A. Whitney\altaffilmark{6}
}
\altaffiltext{1}{Department of Physics and Astronomy, California State Polytechnic University, 3801 West Temple Ave, Pomona, CA 91768, USA; mspovich@cpp.edu}
\altaffiltext{2}{Department of Astronomy and Astrophysics, The Pennsylvania State University, 525 Davey Lab, University Park, PA 16802, USA}
\altaffiltext{3}{Max Planck Institute for Astronomy, K\"{o}nigstul 17, 69117, Heidelberg, Germany}
\altaffiltext{4}{Freelance Consultant, Headingley Enterprise and Arts Centre, Bennett Road Headingley, Leeds LS6 3HN, United Kingdom}
\altaffiltext{5}{School of Physics, University of Exeter, Exeter EX4 4QL, UK}
\altaffiltext{6}{Department of Astronomy, University of Wisconsin--Madison, 475 North Charter Street, Madison, WI, 53706, USA}

\begin{abstract}

We present a catalog of 840 X-ray sources and first results from a 100 ks {\it Chandra X-ray Observatory} imaging study of the filamentary infrared dark cloud G014.225--00.506, which forms the central regions of a larger cloud complex known as the M17 southwest extension (M17 SWex). In addition to the rich population of protostars and young stellar objects with dusty circumstellar disks revealed by {\em Spitzer Space Telescope} archival data, we discover a population of X-ray-emitting, intermediate-mass pre--main-sequence stars (IMPS) that lack infrared excess emission from circumstellar disks. We model the infrared spectral energy distributions of this source population to measure its mass function and place new constraints on the inner dust disk destruction timescales for 2--8~\Msun\ stars. We also place a lower limit on the star formation rate (SFR) and find that it is quite high ($\dot{M}\ge 0.007$~\Msun~\peryr), equivalent to several Orion Nebula Clusters in G14.225--0.506 alone, and likely accelerating. The cloud complex has not produced a population of massive, O-type stars commensurate with its SFR. This absence of very massive (${\ga}20$~\Msun) stars suggests that either (1) M17 SWex is an example of a distributed mode of star formation that will produce a large OB association dominated by intermediate-mass stars but relatively few massive clusters, or (2) the massive cores are still in the process of accreting sufficient mass to form massive clusters hosting O stars.
\end{abstract}


\section{Introduction}
Mid-Infrared surveys of the Galactic plane by the {\em Midcourse Space Experiment} and the {\em Spitzer Space Telescope} have revealed tens of thousands of infrared dark clouds (IRDCs), high-column density, typically filamentary features seen in silhouette against the diffuse 8~\um\ Galactic background emission \citep{S06IRDCs,R06IRDCs}.
IRDCs have been widely studied as the most promising sites providing the initial conditions for massive star cluster formation.
Simulations of collapsing molecular clouds incorporating a variety of physical processes generate filamentary structures at early times over virtually all spatial scales \citep[e.g.][]{VS09,B11,M14,G+VS14}, hence the filamentary nature of dense molecular clouds was not unexpected from a theoretical perspective.
 The {\em Herschel Space Observatory} Hi-Gal survey has confirmed the near-ubiquity of filamentary structures in the Galactic plane, often associated with regions of active, massive star formation \citep{HiGal}.  
These observations have renewed interest in cylindrical accretion geometries \citep{CF53}, ``sausage'' instabilities that form clusters with regular spacings along filaments \citep{N87,Nessie}, and hybrid scenarios involving accretion via cylindrical filaments onto dense, spherical molecular cloud cores, often observed at the nodal points where filaments intersect \citep{M09,H12,PAK13}. 

\begin{figure*}[thbp]
\epsscale{0.95}
\plotone{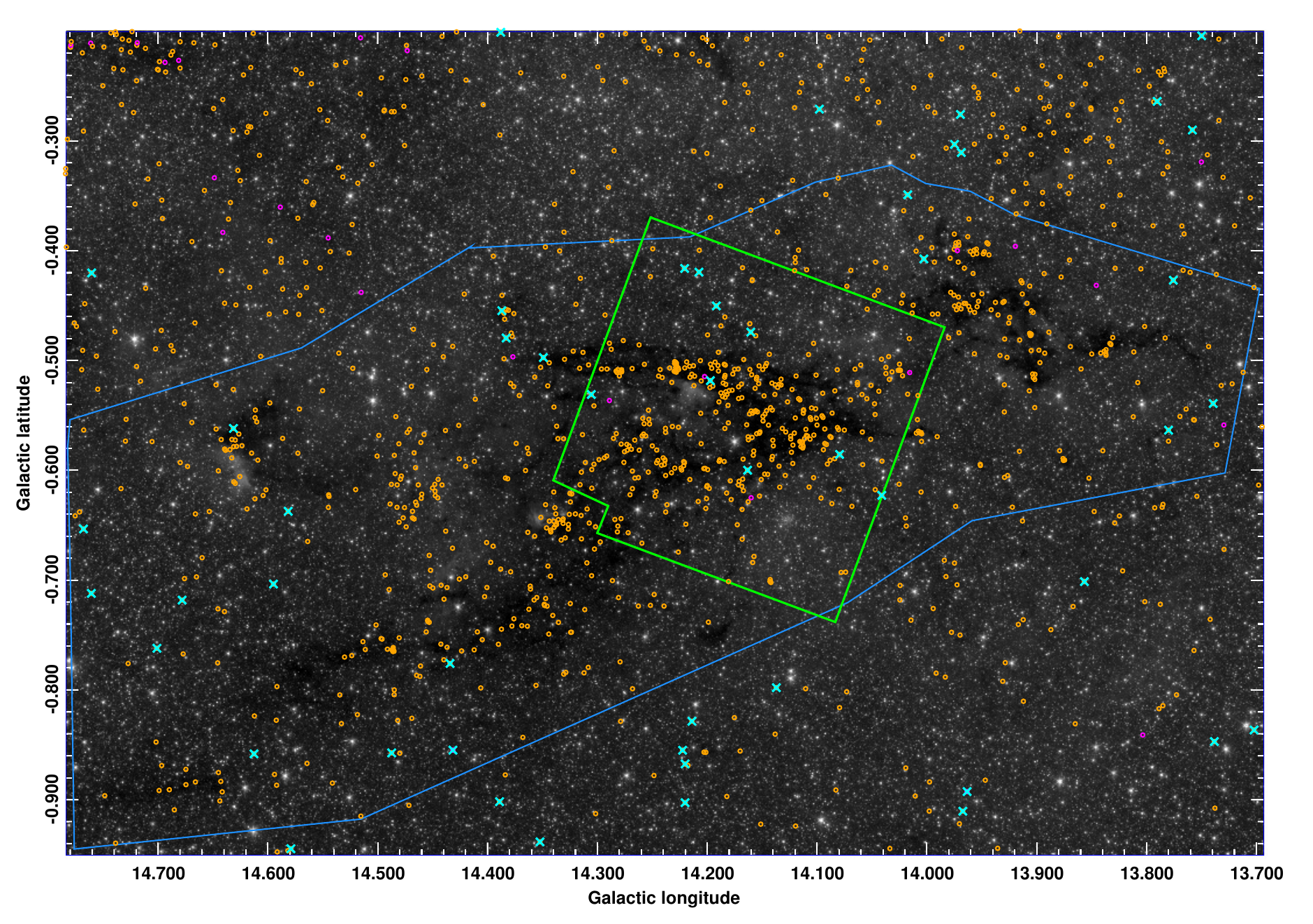}
\plotone{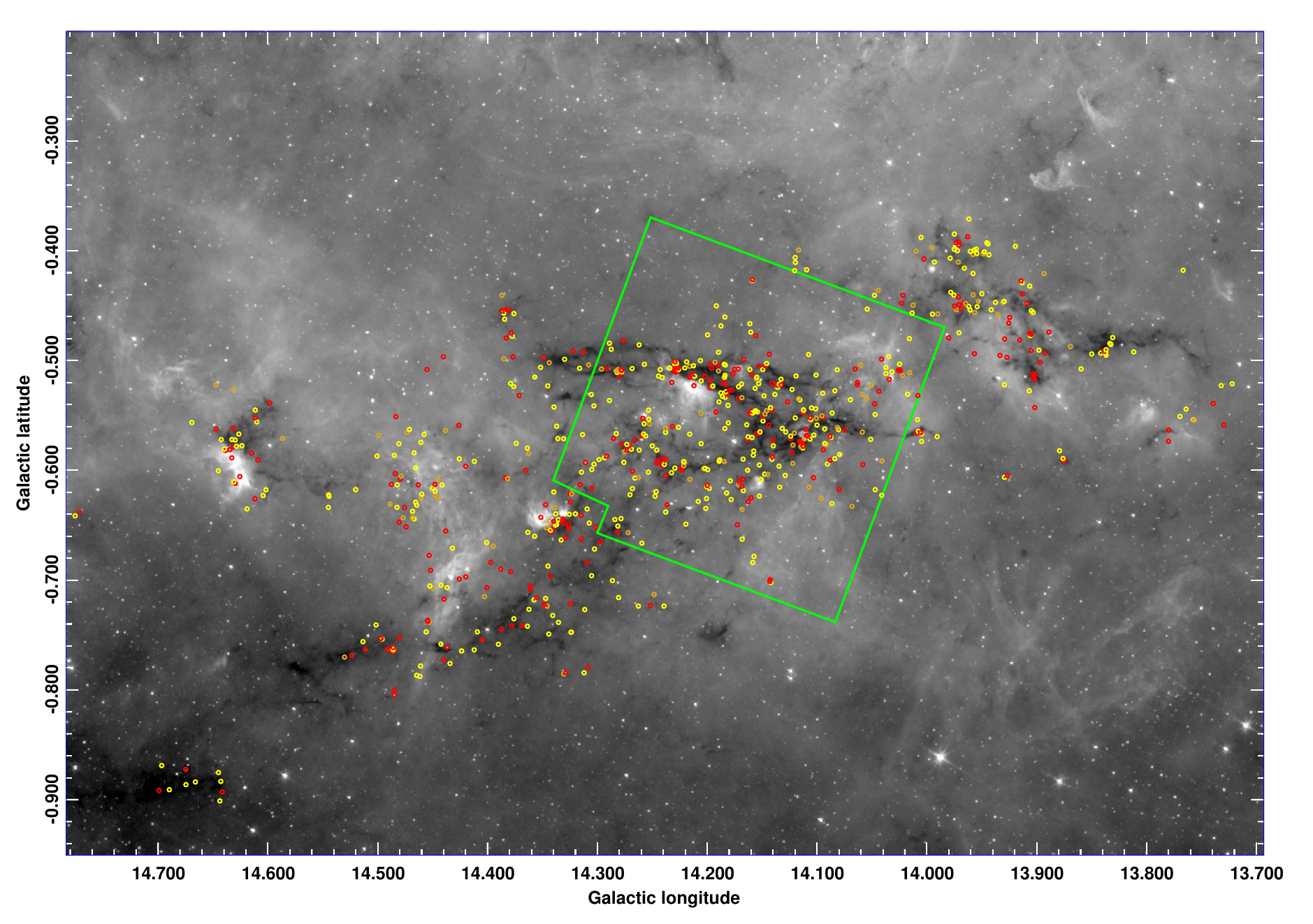}
\caption{IR excess sources in the M17 SWex Wide Field, displayed in Galactic coordinates \citep[for similar visualizations of 18 other star-forming complexes, see][]{MIRES}. 
The M17 H {\scshape ii} region itself is located just beyond the leftmost boundary of this image.
{\it Top:} All IR excess sources overlaid on a GLIMPSE 3.6~\um\ mosaic (orange = candidate YSO, magenta = candidate background galaxy, cyan = candidate dusty AGB star). The green polygon shows the ACIS-I field with one corner removed to exclude the G14.33--0.64 foreground cluster. The blue polygon separates the M17 SWex region from surrounding Galactic ``control'' fields used for clustering analysis. {\it Bottom}: 706 YSOs likely associated with M17 SWex (including possible foreground clouds) overlaid on a GLIMPSE 8.0~\um\ mosaic (red = 253 Stage 0/I YSO, yellow = 353 Stage II/III YSO, orange = 100 ambiguous).
}
\label{fig:ISED}
\end{figure*}

One of the most visually striking IRDC complexes in the Galaxy is the M17 southwest extension, or M17 SWex \citep[][hereafter PW10]{PW10}. 
The morphology of M17 SWex resembles a dark, flying dragon (Figure~\ref{fig:ISED}), extending 1\deg\ (35 pc at $d=2.0$~kpc) to the southwest from the bright M17 \hii region, to which it is connected by a bridge of molecular gas \citep{EL76}. The IRDC G014.225--00.506 comprises the central region of M17 SWex, consisting of a network of dense filaments aligned in two preferred directions, the V-shaped ``dragon's wing.'' This morphology is revealed both in IRDCs and in high-density NH$_3$ molecular gas tracers \citep[][hereafter B13]{B13}. 
The IRDC complex is contained within a larger giant molecular cloud (GMC) with a mass of at least $2\times 10^5$~\Msun\ \citep{ELD79}. In a classic paper proposing a theory for sequential massive star formation triggered by propagating ionization fronts, \citet{EL77} proposed that M17 SWex represents the next generation of massive star formation in the M17 complex, which has already produced dozens of OB stars found in the massive NGC 6618 cluster ionizing the M17 \hii region \citep{HHC97,H08} and in the optically-revealed OB association to the northeast \citep{OI76,P09}.

M17 SWex is, however, already a highly active star-forming region.
Using {\em Spitzer} and Two-Micron All-Sky Survey \citep[2MASS;][]{2MASS} data, PW10 identified 488 predominantly intermediate-mass (2--8~\Msun) young stellar objects (YSOs) via infrared (IR) excess emission from dusty circumstellar disks and protostellar envelopes, representing a total star formation rate (SFR) comparable to that of NGC 6618 itself. 
They also found deficits of OB stars and of massive YSOs compared to the predictions of a standard Salpeter--Kroupa power-law initial mass function \citep[IMF;][]{KW03}. These results suggested that at least some IRDCs can produce rich populations of low- and intermediate-mass stars without commensurate, simultaneous massive star formation. PW10 further predicted that M17 SWex in particular should host a population of young, intermediate-mass pre-main sequence stars (IMPS) that have lost their inner circumstellar dust disks on ${\la}1$~Myr timescales, hence they become indistinguishable from the overwhelming field star contamination in IR surveys of the inner Galactic plane.

Young stars are typically several orders of magnitude more luminous in X-rays compared to older, field stars \citep{PF05}, hence X-ray observations have proven to be a highly effective means of identifying the stellar members of young star-forming complexes in highly contaminated fields, relatively unbiased by the presence or absence of circumstellar disks \citep[e.g.][]{MYStIX,MOXC}. Low-mass T Tauri stars emit X-rays via powerful, convection-driven magnetic reconnection flares \citep{P05} while the strong winds of OB stars produce X-rays through a variety of shock-related mechanisms \citep{F97,G11}. In contrast, intermediate-mass stars, including Herbig Ae/Be stars, have long been regarded as X-ray quiet \citep{S06,CCCPcomps}, as they possess neither strong winds nor strong magnetic fields. There is, however, mounting evidence that the youngest IMPS can be strong, intrinsic X-ray emitters \citep{CCCPYSOs}.

To search for the diskless stellar population predicted by PW10 and hence improve our understanding of the star formation history in M17 SWex, we observed G14.225--0.506 using the Advanced CCD Imaging Spectrometer \citep[ACIS;][]{ACIS} of the {\it Chandra X-ray Observatory}. 
This observation was designed to be combined with near- and mid-IR photometry from the {\it Spitzer Space Telescope} Galactic Legacy Infrared Mid-Plane Survey Extraordinaire \citep[GLIMPSE;][]{GLIMPSE,C09} and MIPSGAL Galactic plane survey \citep{MIPSGAL}, 2MASS, and the UKIDSS Galactic Plane Survey \citep{UKIDSS,UGPS}. 
In Section~\ref{sec:observations} we present the various datasets used for our study. 
In Section~\ref{sec:PCM} we present results of our membership classifications (our classification methodology is described in Appendix~\ref{append:PCMs}). We analyze the spatial distribution and size of the young stellar population in Section~\ref{sec:analysis}. In Section~\ref{sec:discussion} we discuss the implications of our results for star-forming processes in filamentary molecular clouds and prospects for future massive star formation in M17 SWex. Our conclusions are summarized in Section~\ref{sec:conclusions}.

{\em A note on distances.} M17 SWex is kinematically associated with the M17 \hii region and its adjacent GMC, M17 SW, with the bulk of the molecular emission found in the range $v_{\rm LSR}=19$--22~\kms, and the entire complex has historically been assumed to lie at the near kinematic distance of 2.3~kpc (\citealp{EL76,JF82}; B13). Recent VLBI maser parallax measurements have both clarified and complicated the three-dimensional picture of the M17 complex \citep{R14,Wu14}. \citet{S10} reported a distance of $1.1\pm 0.1$~kpc from parallax measurements to the water maser associated with the dense molecular core and ultracompact (UC) \hii region G014.33--00.64, located in the midst of the M17 SWex IRDC lanes. This is highly discrepant with the kinematic distance.
The M17 \hii region itself has an associated methanol maser with a parallax distance of $2.0\pm 0.1$~kpc \citep{X11}. This parallax distance does not strongly depart from the kinematic distance and has become the generally accepted distance to the \hii region and M17 SW. \citet{Wu14} report a parallax distance of $1.8\pm 0.1$~kpc to a water maser associated with G014.63--00.57, located at the end of M17 SWex closest to the M17 \hii region. As \citet{Wu14} note, the parallax distances to G014.63--00.57 and M17 are in agreement, while the significantly nearer distance to G014.33--00.64 indicates that this maser belongs to a foreground cloud. All three masers have similar $v_{\rm LSR}$ and are assigned to the Sagittarius-Carina spiral arm by \citet{R14} and \citet{Wu14}, with M17 and G014.63--00.57 on the far side and G014.33--00.64 on the near side of the arm. 
This raises the possibility that the M17 complex consists of several overlapping clouds along the line of sight.
For the G014.225--00.506 IRDC, the focus of this work, no maser parallaxes are yet available, so we assume $d=2.0$~kpc, the same distance as the M17 \hii region. This choice is motivated by the observation that both the IRDC and the M17 GMC are centered at the same $v_{\rm LSR}=20$~\kms\ (\citealp{P09}; B13), while the G014.33--00.64 maser has multiple velocity components at 19, 22, and 26~\kms\ \citep{S10}. The components at $v_{\rm LSR}=26$~\kms\ are by far the strongest emitters, and this is also the velocity peak of the surrounding, lower-density molecular cloud traced by CO ($J=3\rightarrow 2$) emission observed using the James Clerk Maxwell telescope, part of the highest-resolution CO map obtained to date for the entire M17 complex (C. Beaumont 2012, private communication). This supports the historical picture that the bulk of the M17 complex is at a similar distance, with G014.33--00.64 belonging to a smaller, foreground cloud with a high peculiar velocity.
We discuss the G014.33--00.64 region in more detail in Section \ref{sec:surprising}.

\section{Observations and Basic Datasets}\label{sec:observations}

\subsection{{\chandra} X-ray Observation and Source Catalog}

On 30 and 31 July 2011, we obtained a 98.79 ks continuous \chandra/ACIS exposure centered at $\alpha = 18^h~18^m~17.73^s,~\delta = -16\degree~53\arcmin~48.8\arcsec$ (J2000) using the four detectors of the ACIS-I array plus the S2 and S3 detectors on the ACIS-S array (\chandra\ Observation ID 12348, PI M. S. Povich). ACIS-I has an approximately $17\arcmin \times 17\arcmin$ square field-of-view (FOV) with 0.492\arcsec\ pixels \citep{ACIS}. The point source catalog and analysis presented in this work are restricted to ACIS-I, because the angular resolution of the \chandra\ grazing-incidence mirrors is poor at the large off-axis angles ($\theta > 9\arcmin$) of the ACIS-S detectors, and the ACIS-S detectors covered a part of the sky far from the M17 SWex IRDC filaments. 

Our data reduction, point-source extraction, and photometry procedures have been extensively documented by \citet{AE,CCCPCat} and \citet{MOXC}. We used {\em ACIS Extract} (AE)\footnote{The {\em ACIS Extract} software package and User's Guide are available at \url{http://www.astro.psu.edu/xray/acis/acis_analysis.html}.} \citep{AE,AEcode} to produce a catalog of 840 X-ray point sources detected in our ACIS-I FOV. 
Our point source catalog is published as an electronic FITS table accompanying this article, and its format is identical to the \chandra\ Carina Complex Project (CCCP) X-ray catalog \citep{CCCPCat} and the Massive Star-Forming Regions Omnibus X-ray Catalog \citep[MOXC;][]{MOXC}. 
 
\begin{figure}[tbhp]
\epsscale{1.1}
\plotone{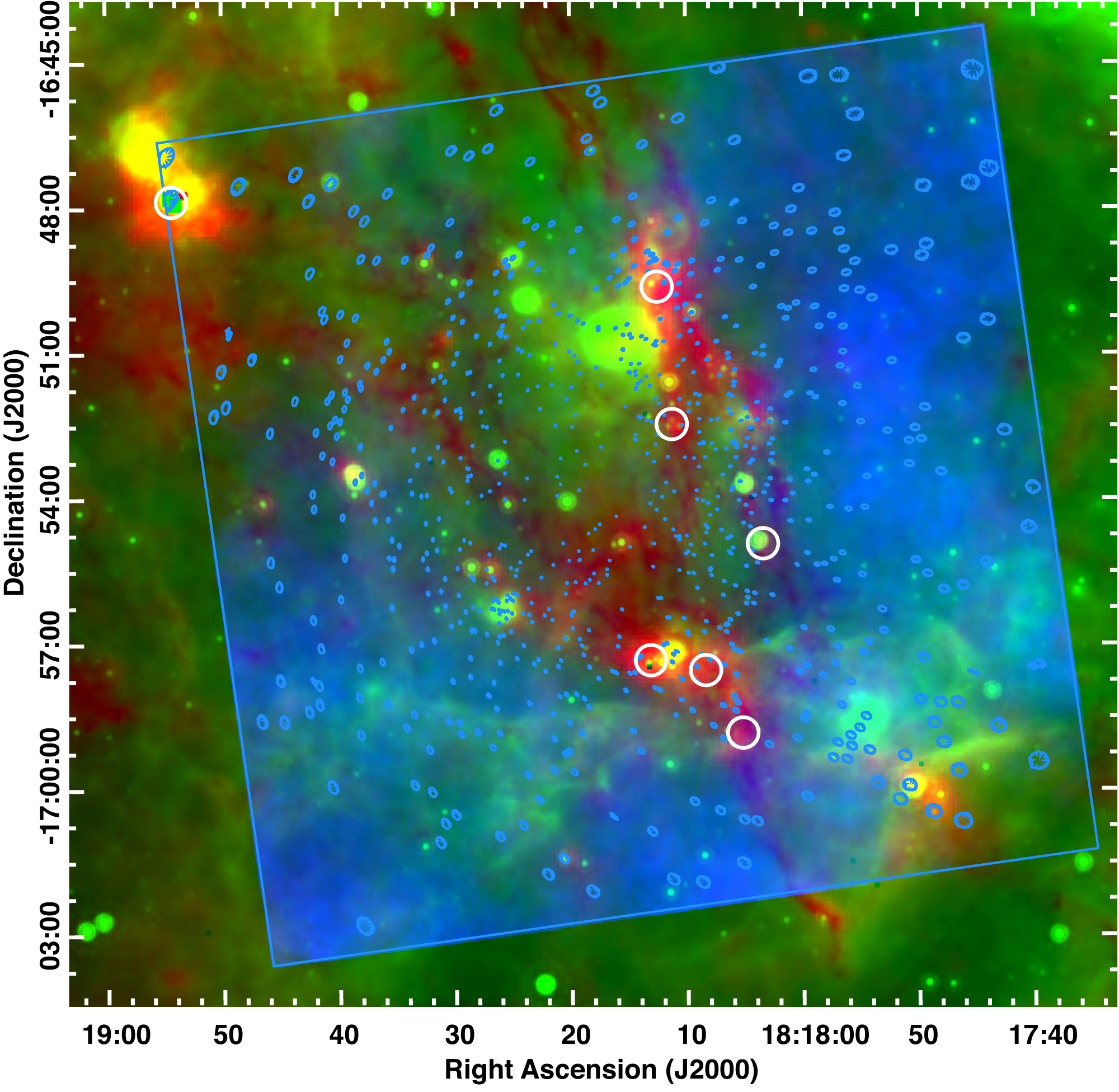}
\caption{Composite X-ray and IR image showing G014.225--00.506, the central region of M17 SWex: red = {\em Herschel} HiGAL 250~\um, green = MIPSGAL 24~\um, blue = ACIS 0.5--2 keV. The 840 X-ray point sources (blue polygons show extraction apertures) have been removed to highlight diffuse X-ray emission. Locations of dense cloud cores from the MALT90 survey \citep{MALT90} 
are marked by white circles.
} 
\label{fig:Xcat}
\end{figure}

A point-source subtracted, soft (0.5--2~keV) X-ray image is combined with IR images of M17 SWex in Figure~\ref{fig:Xcat}, with X-ray point-source extraction apertures overlaid. As Figure~\ref{fig:Xcat} reveals, significant diffuse X-ray emission remains after point sources have been removed; this emission component will be analyzed and discussed in detail in a future paper (Townsley et al., in preparation). 

\subsection{\spitzer\ MIR and {\em 2MASS} NIR Data}

We use mid-infrared (MIR) data at 3.6, 4.5, 5.8, and 8.0 \um\ from GLIMPSE and at 24~\um\ from MIPSGAL, covering the full $1\degree \times 0\fdg 75$ M17 SWex field studied by PW10 (see Figure~\ref{fig:ISED}). This ``IR Wide Field'' spans the entire IRDC complex, and should not be confused with the much smaller ACIS FOV. The GLIMPSE data analysis pipeline produces a highly-reliable point source Catalog that is a subset of a more-complete point source Archive.\footnote{For details of the GLIMPSE pipeline processing and data products, go to \url{http://irsa.ipac.caltech.edu/data/SPITZER/GLIMPSE/doc/glimpse1_dataprod_v2.0.pdf}.} Both the Catalog and Archive incorporate broadband, $JHK_S$ near-IR (NIR) photometry from the {\em 2MASS} Point Source Catalog \citep{2MASS}, which is well-matched to both the 2\arcsec\ resolution of the \spitzer/IRAC detector \citep{IRAC} and the sensitivity limits of GLIMPSE ($[3.6] \la 15.5$ mag, \citealp{C09}). Our standard strategy \citep[see PW10;][]{CCCPYSOs,MIRES} is to use the highly-reliable Catalog to identify IR excess sources produced by the dusty, circumstellar disks and infalling envelopes of young stellar objects (YSOs), and to use the more-complete Archive to provide MIR counterparts to X-ray sources, some of which will identify YSOs that are not present in the more restricted Catalog \citep{MIRES,MYStIXPCM}. Our experience shows that using the Archive instead of the Catalog to search for IR excess sources produces a high false-positive rate, but by using X-ray selection to identify Archive sources of interest we are able to retrieve reliable MIR photometry for sources that are too faint or crowded to appear in the Catalog.
Following PW10 and \citet{CCCPYSOs}, we perform aperture photometry on the MIPSGAL 24~\um\ mosaics {\em only} for IR excess sources identified at shorter wavelengths (see Appendix~\ref{IRE}). This means we do not identify sources that show excess  emission  only at 24~\um, for example transitional disks, because the high potential for source confusion in the inner Galactic plane coupled with the lower, ${\sim}5\arcsec$ resolution of MIPS compared to IRAC and {\em 2MASS}, generates significant numbers of false matches between 24~\um\ sources and field stars \citep{P09}.

\subsection{UKIDSS NIR Photometry}

The combination of unusually high interstellar reddening from the IRDC and circumstellar reddening from protostellar envelopes produces significant differential extinction between NIR and MIR wavelengths. The {\em 2MASS} photometry, used by PW10, is  therefore not deep enough to detect highly reddened stars or deeply embedded protostars detected by GLIMPSE and MIPSGAL in IRDCs such as M17 SWex. We therefore produce deep, $JHK_S$ photometry catalogs by applying the optimal photometry method of \citet{NIRoptimal} to publicly-available UKIDSS data \citep{WFCAM_archive} covering our full M17 SWex field (Figure~\ref{fig:ISED}). The UKIDSS images provide sub-arcsec resolution and point-source sensitivity to $K_S \sim 19$~mag. The UKIDSS spatial resolution approaches the high (on-axis) spatial resolution of our ACIS observation, hence UKIDSS can provide NIR counterparts to close pairs or dense clusters of X-ray sources that are not resolved by \spitzer\ or {\em 2MASS}. We use UKIDSS NIR photometry wherever possible throughout this work, substituting {\em 2MASS} photometry only in cases where the UKIDSS source was saturated or otherwise affected by artifacts. See Appendix~\ref{append:PCMs} for details on how UKIDSS point sources were matched to {\em Spitzer} and {\em Chandra} point sources.

\section{Classification Results: The Combined X-ray and IR Excess Probable Complex Members Sample}\label{sec:PCM}

The results of our classifications for the 840 X-ray catalog sources (described in Appendix \ref{sec:classify}) are presented in Figure~\ref{fig:XSED}. 
\begin{figure*}[thbp]
\centering
\epsscale{1.0}
\plottwo{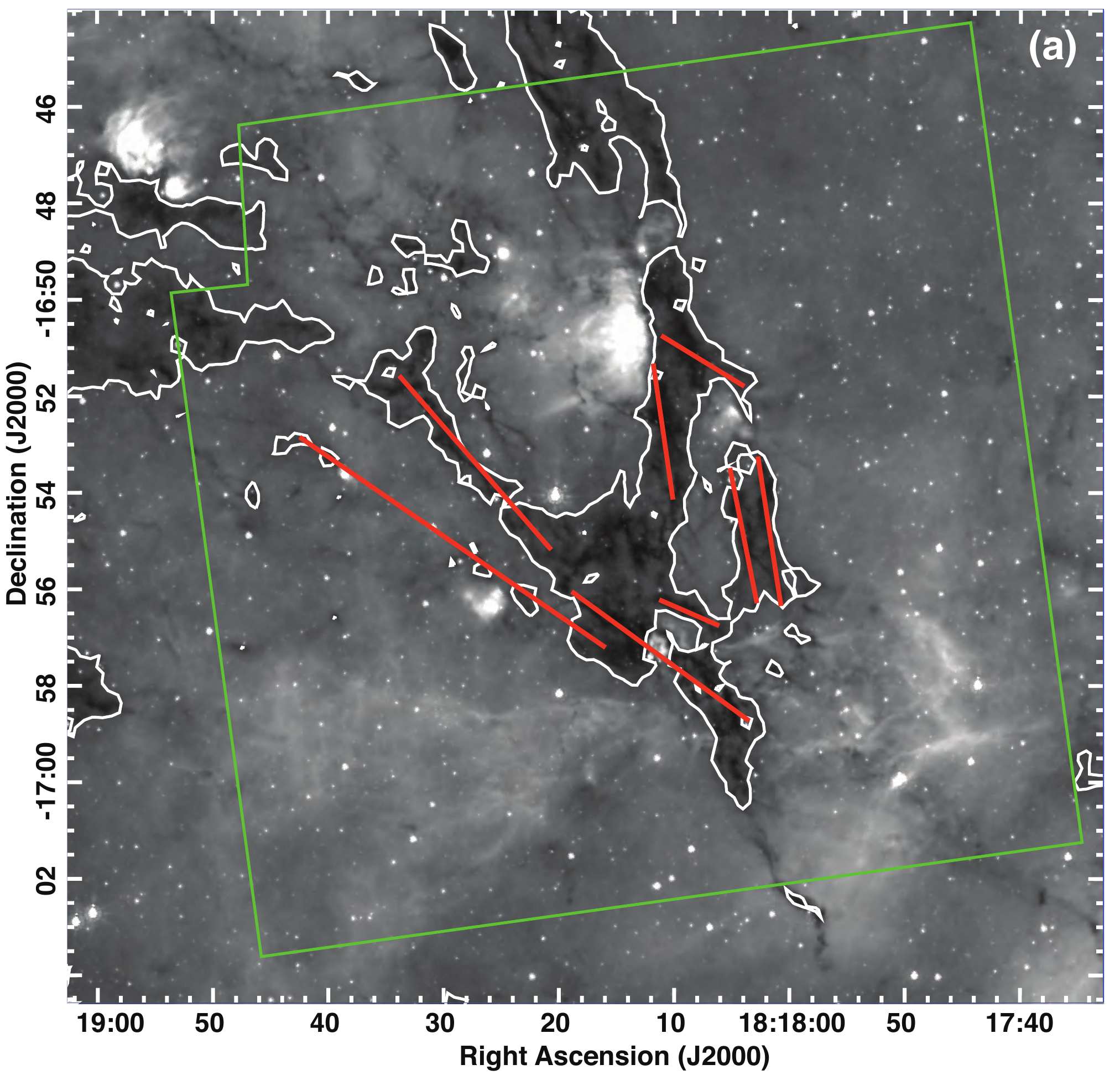}{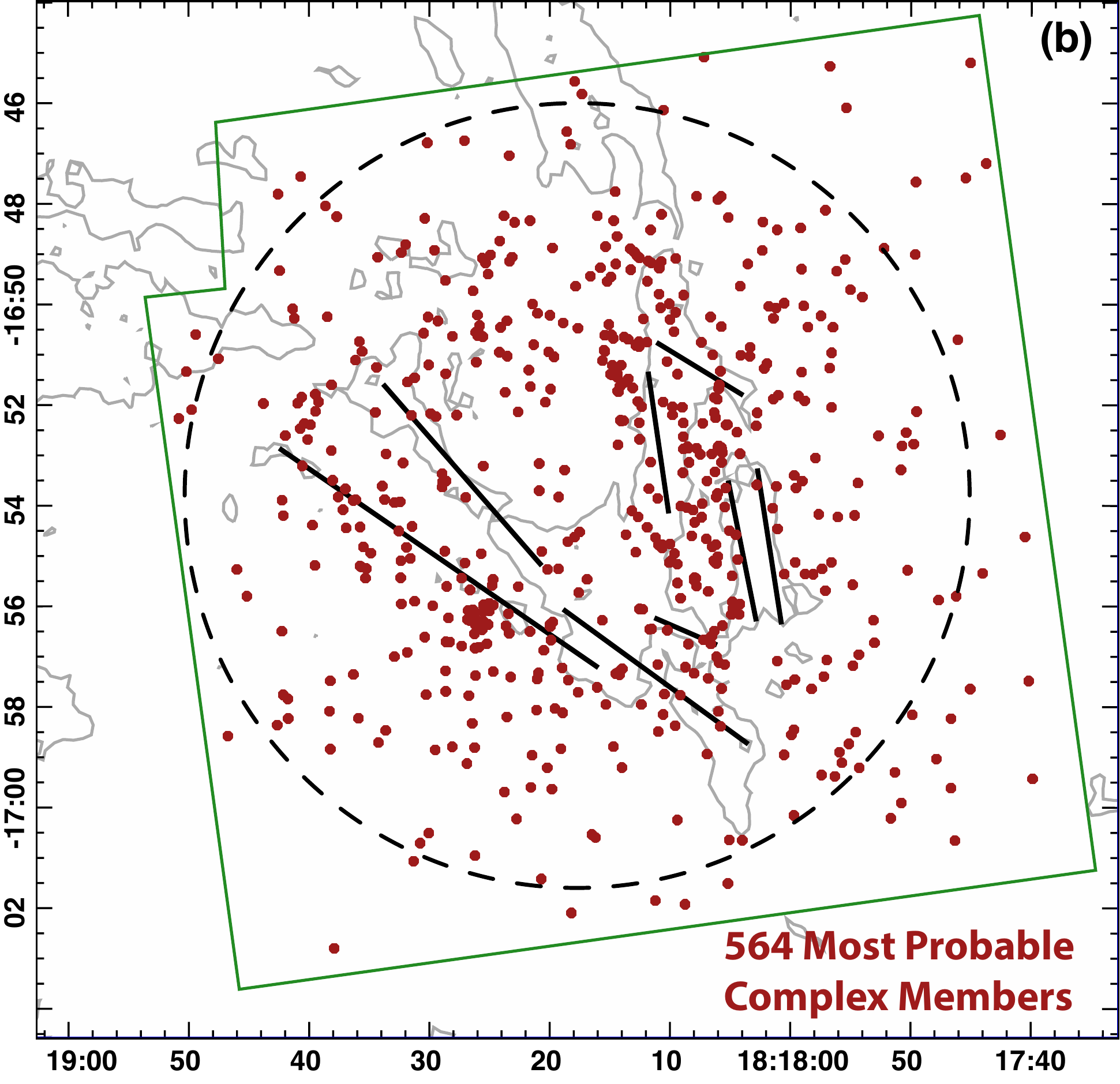}\\
\plottwo{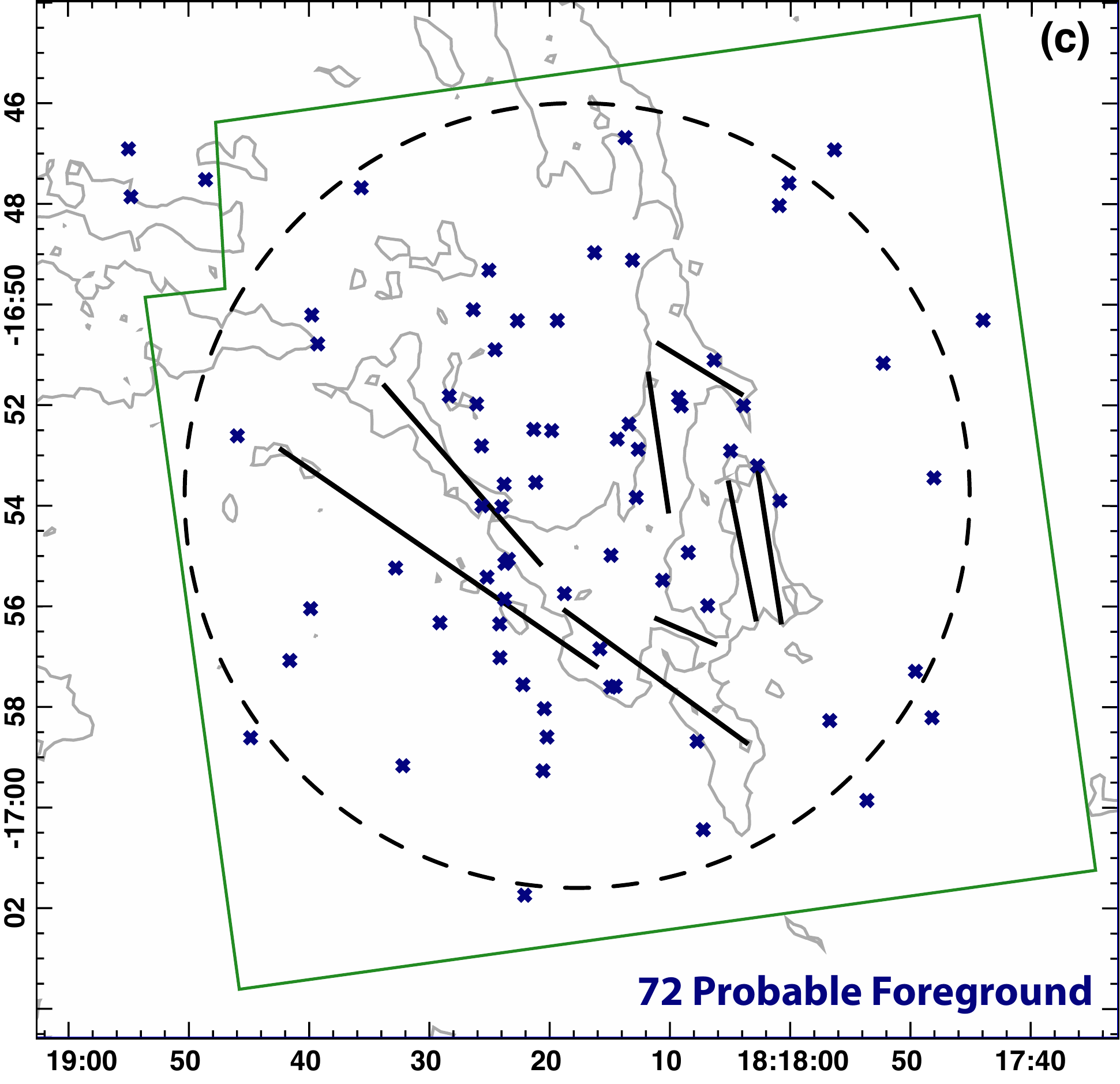}{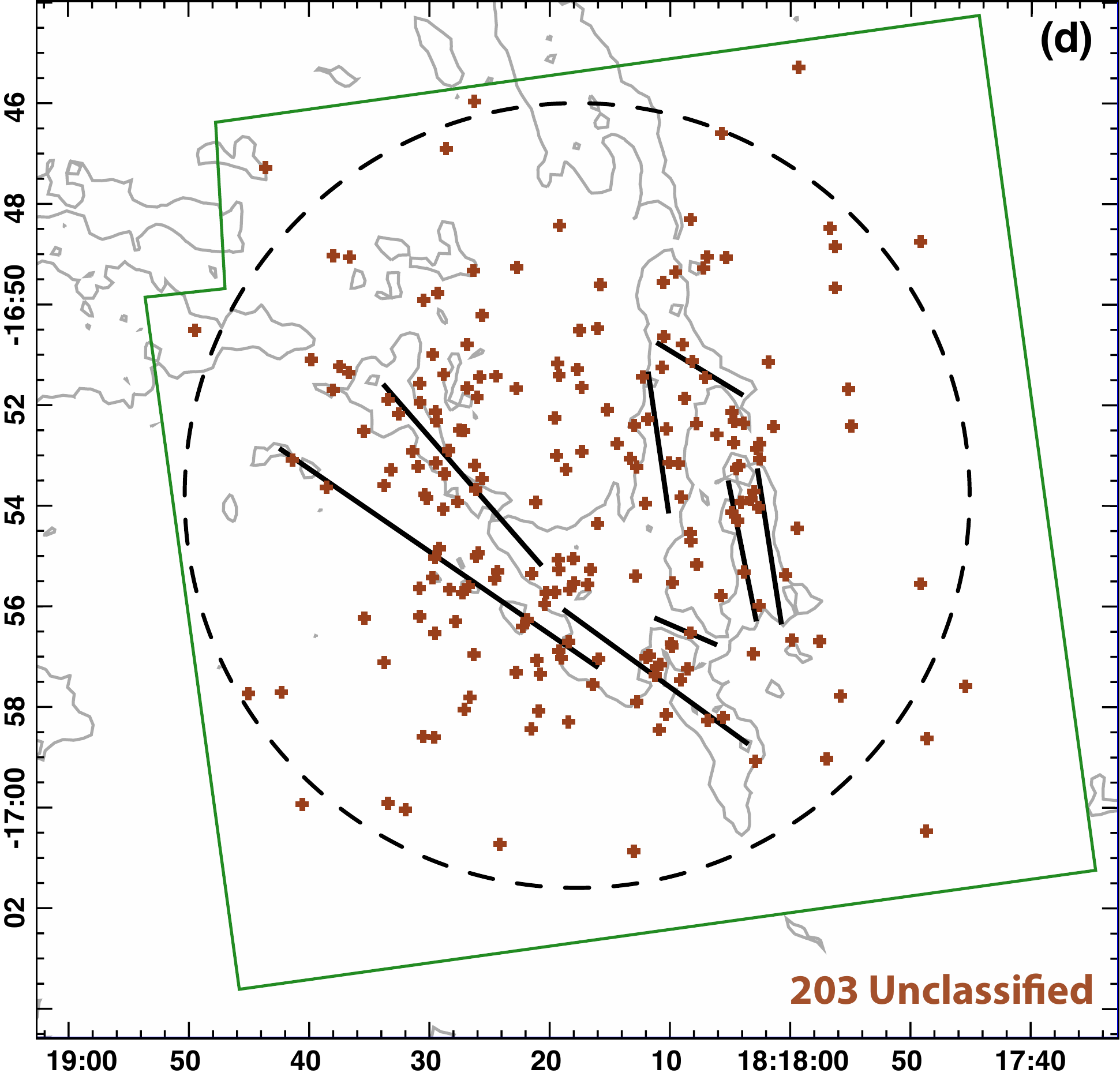}
\caption{(a) IRAC 8~\um\ mosaic overlaid with a (white) contour at 50~MJy~sr$^{-1}$ intensity and (red) line segments corresponding to 8 dense filaments from B13. (b)--(d) Positions of X-ray sources classified as probable members, foreground, and unclassified plotted relative to the same 8~\um\ intensity contour (gray) and filament line segments (black). The dashed circles enclose the area used for filament clustering analysis, $\theta \le 7.8\arcmin$ from the \chandra\ image axis.
In all panels the green polygon outlines the ACIS-I field with one corner removed to exclude the foreground cluster. 
}
\label{fig:XSED}
\end{figure*}
The 564 X-ray-detected probable complex members (PCMs) are generally distributed in a ``V''--shaped pattern that follows the two main IRDC lanes. The 72 foreground stars are more widely distributed across the field, but concentrated more toward the center where the X-ray point-source sensitivity is greater. The spatial distribution of the 203 remaining unclassified sources reflects the strong imprint of the varying X-ray point-source sensitivity, with ${>}85\%$ of unclassified sources having $\theta \le 6\arcmin$, compared to ${\sim}80\%$ of PCM and ${\sim}70\%$ of foreground stars. K--S tests reveal that the foreground, PCM, and unclassified categories all have significantly different distributions with respect to $\theta$, which supports the conclusion that our classification procedure has identified distinct populations of sources with different spatial distributions, correcting for the spatial variations in the X-ray point source sensitivity. 

\subsection{Filament--Imprint Analysis (FIA)}
Spatial distributions that show clustering of sources within obscured star-forming complexes are one of the most powerful lines of evidence that X-ray sources and YSOs trace young stellar populations \citep[e.g.,][]{BFT07,FT08,K08,G09}. Spatial distributions were used as Bayesian priors for the classification of CCCP and MYStIX sources \citep{CCCPClass,MYStIXPCM}. 
Popular techniques for quantifying clustering of young stellar sources include ``minimal spanning trees'' \citep[frequently used in the study of galaxy clustering into filamentary structures, e.g.][]{BBS85,Colberg07} and ``nearest-neighbor analysis,'' a category that includes some types of point-source surface density measurements \citep[e.g.][]{G09}. In this work, we have used Gaussian kernel surface density estimation to identify PCMs detected as YSOs with IR excess emission (Appendix \ref{IRE}, step 7 and Table~\ref{tab:ised}, columns 12 and 13) and nearest-neighbor distances to identify PCMs detected only in X-rays by their proximity to other PCMs (Appendix \ref{sec:PCMprox}).

We can use the spatial distribution of point sources with respect to the observed IRDC filaments to evaluate the results of our X-ray and YSO PCM classifications. If PCMs are preferentially located near the IRDC filaments, this is strong evidence for membership, whereas if they are preferentially located far from the IRDC filaments, or in a distribution that appears uniform (no evident relationship with the filaments), then this calls their membership into question.  The fact that we have already used spatial clustering information in our PCM classifications might seem to bias the results from the start, however the observation that some PCMs cluster near {\em each other} does not presuppose that these groupings should be projected on or near the IRDC.
Furthermore, ${<}10\%$ of the final X-ray PCM sample was identified using spatial clustering (Appendix~\ref{sec:PCMprox}), and an insignificant fraction of YSOs in the ACIS FOV was {\em not} flagged as PCMs on the basis of overdensity (Figure~\ref{fig:ISED}). We need not worry that our spatial analysis will have been biased by the sample selection.

Measuring the distance of point sources from a set of predetermined, linear extended features is a different problem from identifying clusters of point sources or compact extended sources with respect to each other, hence the standard approaches of minimal spanning trees or nearest-neighbor analysis are inappropriate.
We have developed a ``filament--imprint'' analysis (FIA) technique that parameterizes the degree to which the spatial distribution of a given source population bears the imprint of the various linear IRDC filaments in the field.
We represent the locations of the 8 dense filaments identified by B13 as 8 line segments (Figure~\ref{fig:XSED}), each defined by its endpoints in Galactic coordinates.\footnote{Because M17 SWex is located near the Galactic mid-plane, Galactic coordinates provide a convenient, Cartesian coordinate system.}
For the various source samples of interest, we compute the set $s$ of {\em minimum} angular separations between each source and the {\em nearest} ``filament'' line segment (using the IDL routine {\scshape pnt\_line}). We then define
 $\langle s\rangle$
as the {\em median} value of $s$, hence parameterizing the typical separation between that particular source sample and the filaments.  
To simulate a control sample that is not influenced by the presence of the filaments, we run a simple Monte Carlo experiment to generate $n$ uniform samples of coordinates, each with the same {\em average} surface density as the sources in $s$. Then we compute
 $u_i$, the analog of $s$ for {\em each} of the $n$ uniform samples. Finally we define a
dimensionless, ``filament imprint parameter,'' as
\begin{equation}
  f \equiv 1 - \frac{\langle s\rangle n}{\sum^{n}_{i=1}\langle u_i\rangle} = 1 - \frac{\langle s\rangle}{\overline{u}},
\end{equation}
hence $\overline{u}$ is the {\em mean} value of the {\em median} separation $\langle u_i\rangle$ measured across the $n$ uniform samples. 
We have set $n=10,000$ for our simulations.

The logic behind our definition of $f$ is made clearer by considering the astrophysical implications of the limiting cases. $f \rightarrow 1$ for $\langle s\rangle \ll \overline{u}$, in which case the majority of sources lie on or near a linear filament, compared to a random distribution. Astrophysically, this configuration would be approached if the IRDC contained {\em only} YSOs embedded in the centers of the filaments. In general $f>0$ indicates positive correlation between PCMs and filaments. $f \approx 0$ for $\langle s\rangle \approx \overline{u}$, consistent with a random source distribution that bears no imprint of the filaments. Only a foreground population, unaffected by absorption from the IRDC, would be expected to be uncorrelated with the filaments. $f < 0$ is possible for $\langle s\rangle > \overline{u}$, in which case the target source distribution preferentially {\em avoids} the filaments (negative correlation). Thus, astrophysically, negative $f$ values would indicate a background source population strongly absorbed by the IRDC (a sample dominated by AGN, for example).

For the FIA to provide evidence that a given sample of real sources ``knows about'' the IRDC, we must determine whether its measured $f$ departs significantly from our null hypothesis of a uniform configuration. We do this by computing $p_u$, the (two-tailed) statistical $p$-value of our measured $f$ with respect to the simulated distribution produced by replacing $\langle s\rangle$ with $\langle u_i\rangle$ in Equation 1. We consider source samples with $p_u<0.05$ to depart significantly from the null hypothesis, providing convincing evidence that the sources bear the imprint of the IRDC filaments (Table~\ref{tab:FIA}).

For X-ray sources, the FIA is complicated by the variation of point-source sensitivity with off-axis angle $\theta$. We therefore follow the advice of \citet{CCCPCat} and restrict our spatial distribution analysis to the subset of bright X-ray sources that would be detected over the full range of $\theta$ used in the FIA. We find that we must use $\theta \le 7.8\arcmin$, encompassing nearly the entire ACIS FOV (Figure~\ref{fig:XSED}), because for the FIA to work properly we must sample a relatively large area {\em away} from the filaments.
By looking for break points in the power-law distributions of X-ray photon flux $F_{\rm t,photon}$ (or PhotonFlux\_t in the {\em AE} terminology of our X-ray source catalog), stratified by different ranges of $\theta$ \citep[as in Fig.~7 of][]{CCCPCat} we find that our X-ray samples are generally complete for $\log{F_{\rm t,photon}}\ge -6.0$~(photon cm$^{-2}$~s$^{-1}$) and $\theta \le 7.8\arcmin$. These cuts leave 184 X-ray sources (22\% of the full catalog) available for the FIA.

\begin{deluxetable}{lrcc}
\tabletypesize{\small}
\tablewidth{0pt}
\tablecaption{ \label{tab:FIA}
FIA Results for Different Source Samples
}
\tablehead{Sample & size & $f$ & $p_u$\tablenotemark{a}}
\startdata
X-ray PCM  &  158 &  0.35    &  $3\times 10^{-4}$ \\
~~~X-ray diskless\tablenotemark{b} & 83 & 0.28 & 0.04 \\
\hline  
 X-ray foreground    &   12 &  $-0.51$ &  0.12 \\
X-ray unclassified     &   14 &  $-0.78$ &  0.13 \\
\hline
YSOs\tablenotemark{c} &  292  & 0.60 & ${<}10^{-4}$ \\
~~~$A_V^{\rm SED}\le 20$ & 149 & 0.45 & ${<}10^{-4}$ \\
~~~$A_V^{\rm SED} > 20$  & 143  & 0.72 & ${<}10^{-4}$ \\
\enddata
\tablenotetext{a}{We adopt $p_u<0.05$ as the threshold where $f$ (whether positive or negative) departs significantly from the null hypothesis of a uniform source distribution.}
\tablenotetext{b}{Subset of X-ray PCMs with non- or marginal-IR excess counterparts fit by naked stellar SED models.}
\tablenotetext{c}{All candidate YSOs classified as PCM within $\theta\le 7.8\arcmin$ of the \chandra\ pointing, regardless of X-ray detection.}
\end{deluxetable}

\subsection{FIA Results}
The FIA results, $f$ and $p_{u}$ for various source samples of interest, are presented in Table~\ref{tab:FIA}.
The sample size in the second column reflects the number of sources within $\theta\le 7.8\arcmin$, and for samples of X-ray selected sources this number is further reduced by our $F_{\rm t,photon}$ cut to remove faint sources.
Because of these differing selection criteria, FIA would not produce meaningful results for the combined PCM sample of X-ray and IR excess-selected sources.

The 158 X-ray PCMs used in the FIA (33\% of all X-ray PCMs) show a significant imprint of the filaments ($f=0.35$ with $p_u=3\times 10^{-4}$; Table~\ref{tab:FIA}). The 292 YSOs used in the FIA exhibit a higher degree of clustering ($f=0.60$ with $p_u<10^{-4}$) compared to X-ray PCMs, which include both YSOs and non-IR excess sources. These results support our conclusion that the PCMs are indeed associated with M17 SWex. For the foreground and unclassified sources, we find only a handful ($n=12$ and 14, respectively) that meet the requirements for the FIA. To the degree that we can estimate $f$ for such small numbers of sources, we find {\em negative} $f$ values, suggesting an tendency to avoid the IRDC filaments, but the $p_u$-values are consistent with the null hypothesis of a uniform source density.

We assume implicitly that the bright X-ray sources used for FIA are representative of the spatial distribution of the full population. This assumption may not be valid, but we are comfortable making it because the fainter sources include many PCMs that are highly-obscured by dust in the IRDC lanes. Hence we expect a bias toward {\em higher} filamentary clustering of fainter PCMs, which would suggest that our $f$ values obtained for the brighter sources {\em underestimate} the true degree of clustering among the PCMs. We see precisely this effect when applying FIA to the YSOs stratified by \avsed; while both samples are significantly clustered, YSOs with $A_V^{\rm SED} > 20$~mag are {\em more} clustered, and at a higher significance level, compared to the less-obscured YSOs (Table~\ref{tab:FIA}, and we further discuss this effect in Section~\ref{sec:spatial} below).

We conclude that our combined IR-excess and X-ray source classification methodology, presented in Appendix~\ref{append:PCMs},
works well to identify young stellar populations provided (1) there is sufficient foreground extinction toward PCMs ($A_V>2$~mag) to allow separation of foreground stars from PCMs, and (2) high background absorption screens out the majority of background stars and AGN. Any Galactic IRDC should fit these conditions, and our FIA results show that they hold for M17 SWex. The spatial distribution of less-obscured foreground sources does not bear an imprint of the IRDC. A significant fraction of mis-classified background sources among the X-ray PCMs would dilute or erase the observed clustering along the filaments.

\section{Analysis of the PCM Population}\label{sec:analysis}

\subsection{Spatial Distribution and Imprint of Disk Evolution}\label{sec:spatial}
\begin{figure}[htp]
\epsscale{1.1}
\plotone{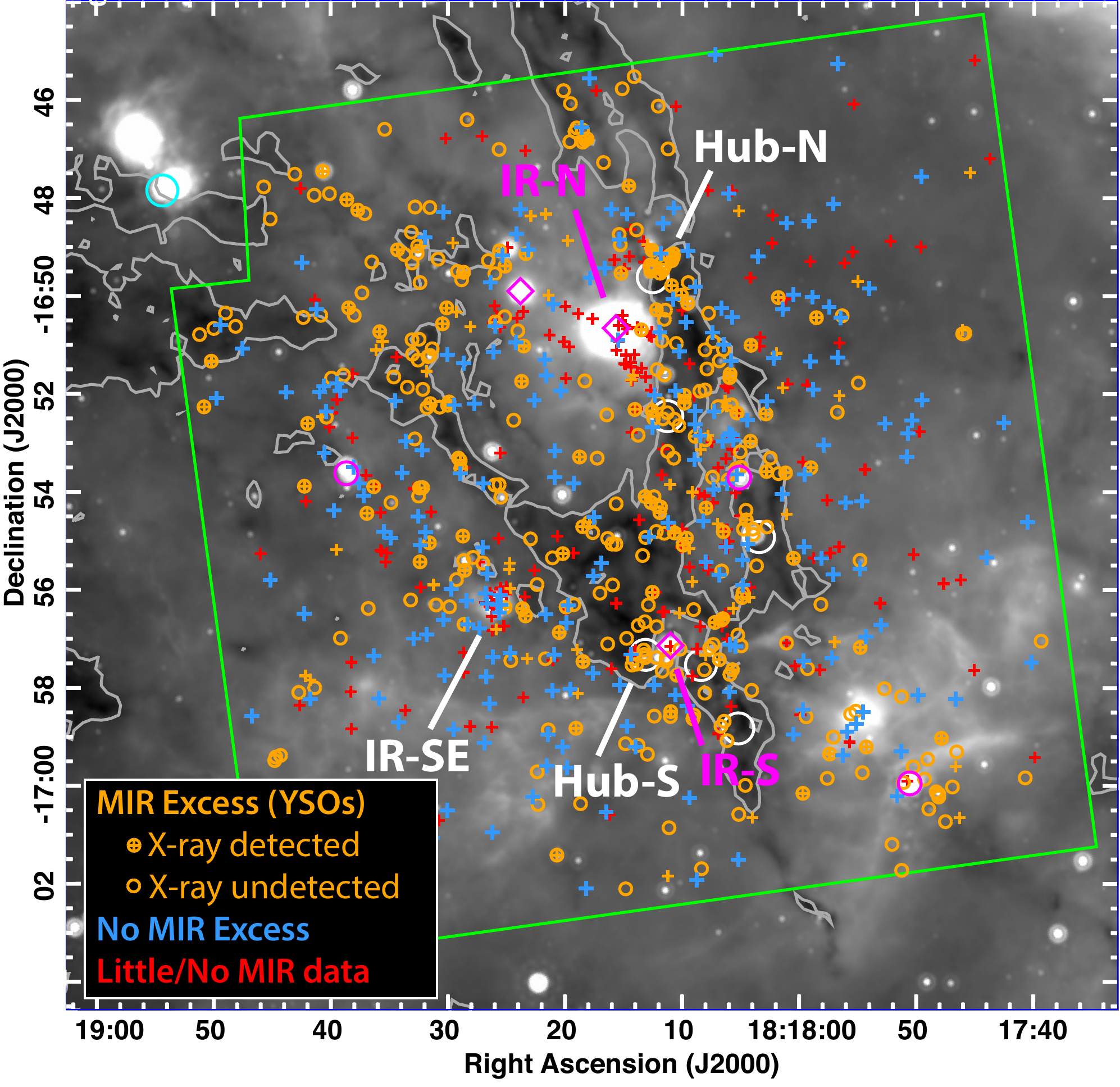}
\caption{
All PCMs in the ACIS FOV, including stars identified via X-ray emission (crosses) and YSOs identified by IR excess (circles; YSOs detected in X-rays are circles around crosses) plotted on a MIPSGAL 24~\um\ image. Symbol colors indicate presence or absense of MIR excess emission: yes (orange), no (blue), or undetermined
 (red). Three larger, magenta circles mark candidate massive YSOs detected by \msx\ (PW10), and 3 magenta diamonds mark other bright regions of 24~\um\ emission likely powered by early B stars not included in our PCM sample. Large, white circles of 38\arcsec diameter show the MALT90 cores listed in Table~\ref{tab:MALT90} (the cyan circle at the top left is the MALT90 core associated with the probable foreground cloud G14.33-0.64). The 8~\um\ contour from Figure~\ref{fig:XSED} outlines the darkest regions of the IRDC.
}
\label{fig:members}
\end{figure}

Our final, combined sample consists of 847 PCMs in the ACIS FOV, broken down as 510 stars and YSOs identified via X-ray emission, 236 YSOs identified by IR excess, 98 YSOs identified both by IR excess and as X-ray sources, and 3 luminous YSOs, detected by \msx, from the PW10 catalog. As shown in Figure~\ref{fig:members}, no single cluster dominates the source distribution, instead the sources are generally concentrated near the IRDC filaments, as we found using FIA, with subclustering on smaller scales.  
The most prominent, small subclusters visible in Figure~\ref{fig:members} are all located in or near several compact, bright 24~\um\ nebulae or 6 dense molecular cores in the ACIS field cataloged by the MALT90 survey \citep{MALT90}. We compile basic properties of the MALT90 cores in Table~\ref{tab:MALT90}. 
For all cores, we use the isothermal dust model of \citet{H13} to calculate the temperature $T_d$, molecular hydrogen column density $N_{H_2}$, and mass $M_{\rm kin}$ assuming the individual kinematic distance $d_{\rm kin}$ to each core. We also report $M_{2.0}$, which scales $M_{\rm kin}$ to our assumed distance of $d=2.0$~kpc (column density, and hence mass, is directly proportional to IR intensity, so the masses scale as the inverse-square of the distance).

\begin{deluxetable*}{lccccccc}
\tabletypesize{\small}
\tablewidth{0pt}
\tablecaption{ \label{tab:MALT90}
Properties of MALT90 Cores within the M17 SWex ACIS Field
}
\tablehead{
 & & \multicolumn{4}{c}{Isothermal dust model} & B16\tablenotemark{f} & \\
\cline{3-6}
 & $d_{\rm kin}$\tablenotemark{a} & $T_d$\tablenotemark{b} & $N_{H_2}$\tablenotemark{b} & $M_{\rm kin}$)\tablenotemark{b} & $M_{2.0}$ & $M_{\rm tot}$ & Associated \\
           Core name       & (kpc)    & (K) & ($\times 10^{21}$~cm$^{-2}$) & (\Msun)  &   (\Msun) &  (\Msun) & Sources\tablenotemark{c} 
}
\startdata
G014.077--00.558 & 2.59 & 19 & 9.1 & 65 & 39 & & 24 \\
G014.102--00.559 & 2.51 & 22 & 9.8 & 73 & 46 & & X,Y \\
G014.114--00.574\tablenotemark{d} & 2.37 & 28 & 10  & 68 & 48 & 717 & X,Y,24,C \\ 
G014.131--00.521 & 2.44 & 15 & 6.4 & 45 & 30 & & Y,C \\
G014.182--00.529 & 2.43 & 14 & 2.1 & 13 & 9 & & Y,C \\
G014.226--00.511\tablenotemark{d} & 2.24 & 27 & 19  & 107 & 85 & 979 & X,Y,24,C \\
\hline
G014.331--00.644 & 2.55\tablenotemark{e} & 23 & 57 & 440 & 82\tablenotemark{e} & & X,Y,C \\
\enddata
\tablenotetext{a}{Near kinematic distance of MALT90 source \citep{MALT90}.} 
\tablenotetext{b}{Dust temperatures, column densities, and masses were calculated using {\em Herschel} Hi-Gal data, as described in \citet{H13}.}
\tablenotetext{c}{``Associated sources'' are those projected within a 38\arcsec\ Mopra beamwidth centered on the MALT90 core (Figure~\ref{fig:members}), and are described by: X = X-ray source(s), Y = YSO(s), 24 = MIPS~24~\um\ source(s) not detected at shorter wavelengths, C = ``cluster'' of three or more X/Y sources.}
\tablenotetext{d}{MALT90 sources G014.114--00.574 and G014.226--00.511 correspond to NH$_3$ filament Hub-S and Hub-N, respectively (B13).}
\tablenotetext{e}{For this core we assuming a (foreground) distance of $d=1.1$~kpc determined from trigonometric parallax measurements of the associated maser \citep{S10}, hence the corrected mass reported is $M_{1.1}$. All other corrected masses assume $d=2.0$~kpc.}
\tablenotetext{f}{Values reported by \citet{B16} for Hub-S and Hub-N only.}
\end{deluxetable*}

The two filament nodes, Hub-N and Hub-S (B13), correspond to the two most massive MALT90 cores, G014.226--00.511 and G014.114--00.574, respectively (Table~\ref{tab:MALT90}), both of which have associated YSO subclusters but relatively few X-ray sources (Figure~\ref{fig:members}). The masses inferred by B13 from NH$_3$, a dense gas tracer, are ${\sim}200$~\Msun\ and ${\sim}250$~\Msun\ for Hub-N and Hub-S, respectively (assuming $d=2$~kpc). \citet{B16} simultaneously fit the combined Hi-Gal plus sub-mm SEDs and the sub-mm radial intensity profiles and derived total masses of $717\pm 250$~\Msun\ and $979\pm 329$~\Msun\ for Hub-S and Hub-N, respectively (also assuming $d=2.0$~kpc).
These total dust masses are higher than the corresponding isothermal dust masses we derived from Hi-Gal emission by factors of ${\sim}15$ and ${\sim}11.5$ for Hub-S and Hub-N, respectively. The masses reported by \citet{B16} should be the most reliable available, because the inclusion of sub-mm data in the SED measures the contribution from the large reservoir of very cold dust surrounding the core, and their assumption of a radial dependence of dust temperature is more realistic than an isothermal core. We therefore suspect that the masses for the other MALT90 cores listed in Table~\ref{tab:MALT90} may similarly be underestimated by an order of magnitude.

\begin{figure*}[thbp]
\epsscale{1.0}
\plottwo{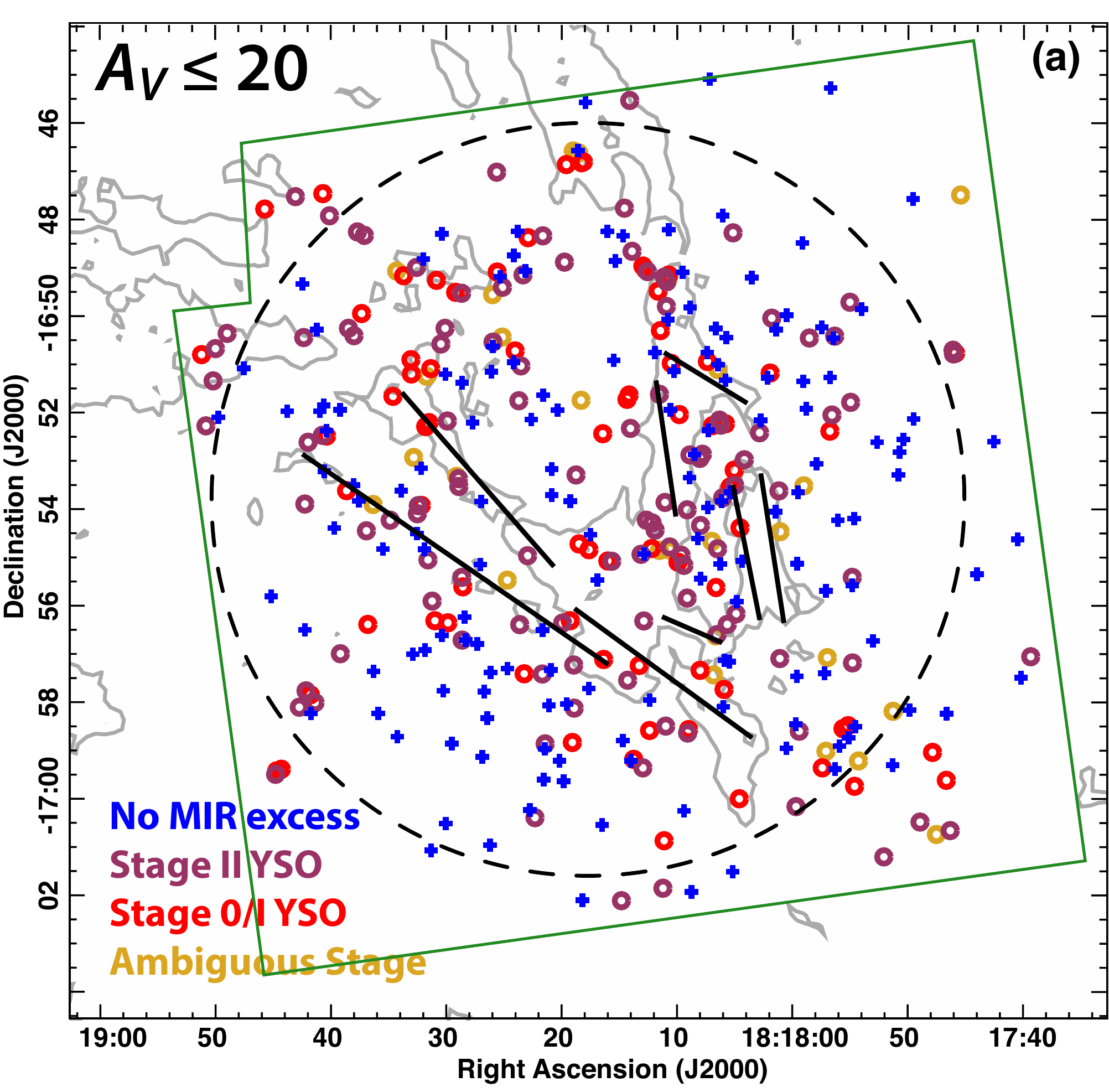}{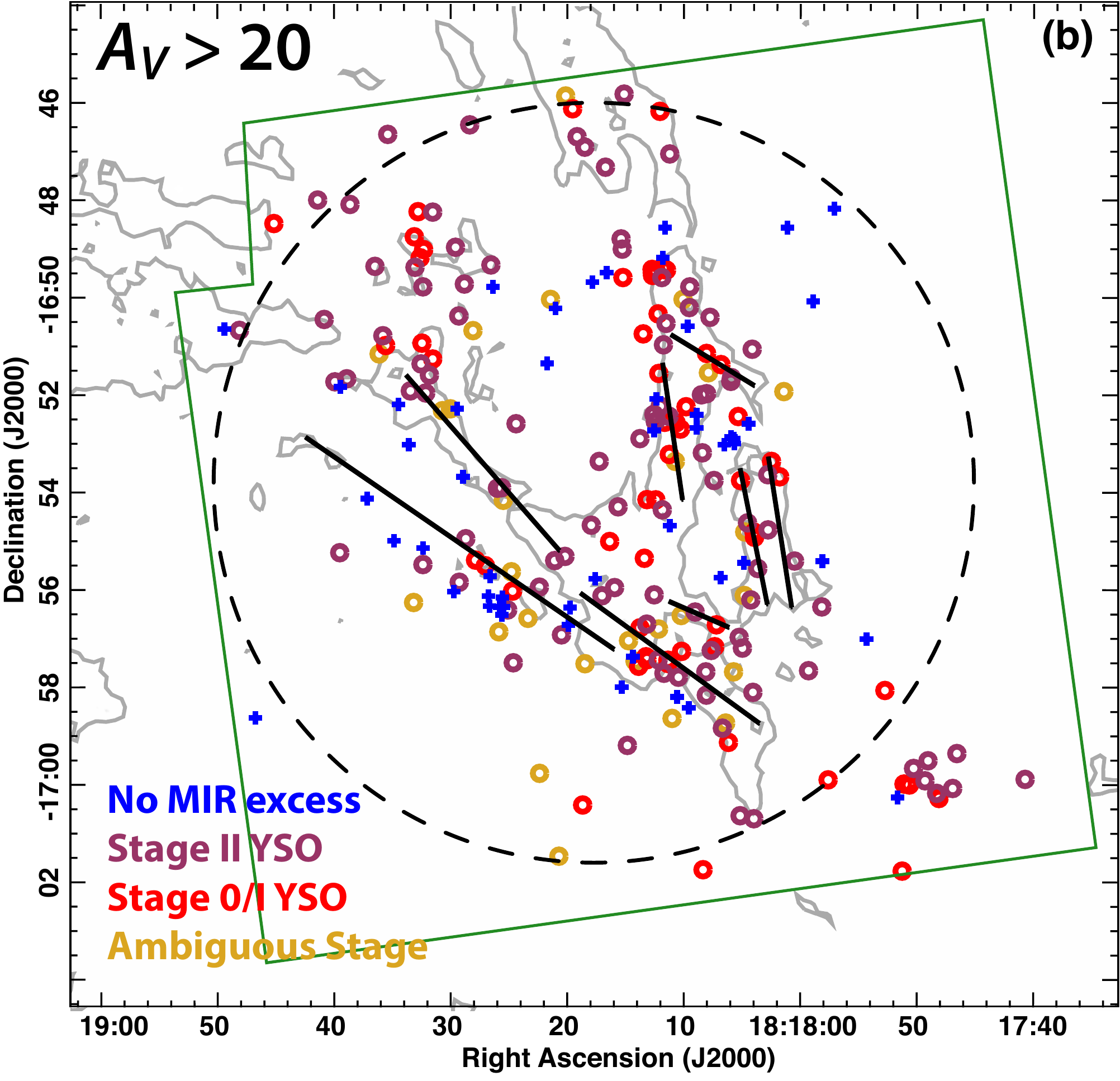}
\caption{Spatial distributions of all PCMs with (a) $A_V\le 20$~mag and (b) $A_V > 20$~mag determined from SED fitting. Crosses mark X-ray selected stars without MIR excess emission and circles mark YSOs irrespective of X-ray detections. Other overlays and contours are the same as in Figure~\ref{fig:XSED}.
}
\label{fig:avstrat}
\end{figure*}
We correspondingly designate three bright, resolved \spitzer\ MIR sources in the ACIS  FOV as IR-N, IR-S, and IR-SE (Figure~\ref{fig:members}; all are {\em IRAS} point sources noted by B13). IR-N and IR-S are adjacent (in projection) to Hub-N and Hub-S, respectively, while IR-SE is located in a dense filament extending NE from Hub-S, the {\em only} filament identified by B13 that is {\em not} traced by an IRDC over its full length (see Figures \ref{fig:XSED}--\ref{fig:avstrat}). In contrast to the hubs, each of which has an associated cluster of protostars, IR-N and IR-SE each contain one of the two richest clusters of X-ray sources detected in M17 SWex. 

The majority of the ${\sim}25$ X-ray sources associated with IR-N have neither a MIR counterpart for SED analysis (the bright MIR emission of IR-N inhibited GLIMPSE point-source detection) nor UKIDSS $J$ detections to enable NIR excess determinations (the cluster is highly obscured). There is a MIR-detected star at the center of IR-N, undetected in X-rays and with no IR excess, which may be the ionizing star responsible for lighting up the MIR nebula. The IR and radio continuum luminosity of IR-N \citep{JF82,RMS11a} correspond to a B1-1.5 V star. 

A similar number of X-ray sources (21) are detected in IR-SE as in IR-N, but they are more tightly clustered. Eight have SED fitting results, of which two are YSOs and the remaining six show no evidence for significant IR excess. The most luminous of these non-IR excess sources is the IR counterpart to the X-ray source CXO J181825.47-165618.4, which was detected in four bands: $H=16.73$, $K_s=14.007$, $[3.6]=11.88$, and $[4.5]=11.11$~mag. This very red IR spectral index implies very high foreground extinction, and the SED fitting results using our ``naked'' stellar models (Appendix~\ref{append:construct}) return a luminosity equivalent to a very early B-type star, with $A_V^{\rm SED}=39$~mag. The bolometric luminosity of this one star, reprocessed by dust, is more than sufficient to power IR-SE, which is not as bright as IR-N. The associated X-ray source was detected with 8 net counts, primarily in the hard band (2--8 keV, $E_{\rm med}=3$~keV) with no evidence for variability. Preliminary X-ray spectral fitting using XSPEC \citep{XSPEC} is also consistent with a highly-obscured, massive star (total hydrogen column density is
$\log{N_H} = 22.7$, and absorption-corrected 0.5--8~keV X-ray luminosity is $\log{L_{t,c}}        =   30.4$). 

We detect six X-ray sources apparently associated with IR-S, only one of which, CXO J181810.98--165708.9, was classified as a PCM on the basis of an IR counterpart detected at 8~\um\ only in the GLIMPSE Archive (and visual inspection of the 8~\um\ images casts doubt on the reliability of this 8~\um\ source). The remaining five sources remain unclassified because they all lack IR counterparts and were too far separated from PCMs to trigger our proximity classification (Appendix~\ref{sec:PCMprox}). All six sources have $E_{\rm med}>3$~keV, indicating very high absorption. The brightest of these, CXO J181811.29-165722.2, was detected with 28 net counts, {\em all} in the hard band ($E_{\rm med}=4.15$~keV), with no evidence for variability. 
Preliminary XSPEC fitting results for this source indicate a highly-obscured, massive star ($\log{N_H} = 23.2$, $\log{L_{t,c}}       =   31.2$).
The IR and radio luminosity of IR-S correspond to a B1.5-2.5 V star \citep{JF82,RMS11a}. 
In a subsequent paper we will present, via X-ray spectral fitting, the detailed properties of the X-ray source population in M17 SWex.

Both IR-N and IR-S are detected in radio continuum \citep{JF82}, which places an indirect lower bound on their distances. If these sources were at the 1.1~kpc distance of the \citet{S10} maser, the IR luminosities shift downward by the equivalent of several spectral subclasses: IR-N would become a B2.5 V star and IR-S would be later than a B3 V star. Because ionizing luminosity is a stronger function of spectral type than bolometric luminosity, the spectral types at 1.1~kpc, particularly for IR-S, are incompatible with their observed radio continuum luminosities.

The six stars associated with IR-SE that lack MIR excess emission are part of a much larger PCM subpopulation consisting of 209 ``diskless'' X-ray sources with MIR counterpart SEDs indicating an absence of dusty inner disks (blue crosses in Figure~\ref{fig:members}). The existence of this population was predicted by PW10, who observed a deficit of YSOs at higher luminosities and suggested that rapid disk evolution among IMPS could be responsible. We would like to compare the spatial distributions of the diskless stars to the YSOs to search for possible evolutionary trends. To make this comparison, we first divide the PCMs into two subsamples: less-obscured with $A_V^{\rm SED} \le 20$~mag (Figure~\ref{fig:avstrat}a) and heavily-obscured with $A_V^{\rm SED} > 20$~mag (Figure~\ref{fig:avstrat}b). The dividing \avsed\ value was chosen to separate the principal locus of X-ray selected PCMs on the NIR CMD from the highly-incomplete sample of more-reddened sources (Figure~\ref{fig:JHK}b). 
The 83 X-ray diskless stars to which we can apply our FIA show marginally significant correlation with the filaments ($f=0.28$ with $p_u=0.04$; Table~\ref{tab:FIA}). The majority of these (73) are in the less-obscured population ($A_V^{\rm SED}\le 20$~mag), and the corresponding sample of 149 less-obscured YSOs is more strongly correlated with the filaments ($f=0.45$ with $p_u<10^{-4}$).

The X-ray detected sources with no MIR excess in the heavily-obscured population are too few and too faint to provide reliable FIA results, but it is clear from Figure~\ref{fig:avstrat}b that they are concentrated in the IRDC filaments, along with the heavily-obscured YSOs (which exhibit the greatest clustering along the filaments; see Table~\ref{tab:FIA}).

The outlines of a 3-dimensional star formation history reflecting the filamentary IRDC geometry emerge from Figure~\ref{fig:avstrat}. The less-obscured PCMs are distributed throughout an extended halo of lower-density molecular gas surrounding the IRDC filaments \citep[as observed toward other IRDCs, see][]{IRDC_NIR}. Moving deeper toward the centers of the IRDC filaments, where $A_V> 50$~mag \citep{BT09}, we find the more-obscured PCMs clustered along the filaments. This spatial distribution is accompanied by an apparent spread in stellar ages. The diskless X-ray PCMs represent a more evolved, less-obscured, and less-clustered (with respect to the filaments) population compared to the YSOs. The apparent dearth of diskless stars among the heavily-obscured sample may be biased by selection effects against detecting non-IR excess, X-ray sources deeply embedded within the IRDC, but the significant difference in $f$ when comparing less-obscured diskless stars and YSOs appears to be a real, evolutionary effect. We can conclude that the PCM population in M17 SWex exhibits a large-scale ``filament-halo'' age gradient analogous to the core-halo age gradients recently reported in the Orion Nebula and NGC 2024 massive clusters \citep{GFK14}. 

Given that protostars, disk-dominated pre-MS stars (including classical T Tauri stars), and finally stars lacking inner disks (including weak-lined T Tauri stars) represent a progression along the same evolutionary sequence, we might expect to see signs of a filament-halo age gradient when comparing Stage 0/I to Stage II YSOs.
In Figure~\ref{fig:avstrat} we identify YSOs at different evolutionary stages.  
We find no significant differences in $f$ with respect to YSO evolutionary stage, hence we do not report separate FIA results for Stage 0/I and Stage II YSOs in Table~\ref{tab:FIA}, with the caveat that our stage classifications are often uncertain (a main reason we include an ambiguous classification for YSOs with poorly-constrained model fits). In the less-obscured sample  (Figure~\ref{fig:avstrat}a), the Stage 0/I and Stage II YSOs appear well-mixed spatially, but in the highly-obscured sample (Figure~\ref{fig:avstrat}b) the Stage 0/I protostars appear more concentrated in dense clusters, while the Stage II YSOs appear more distributed along the filaments. This general pattern is observed across the entire M17 SWex cloud (Figure~\ref{fig:ISED}b) and in numerous other Galactic molecular clouds spanning a wide range of distances and masses \citep[e.g.][]{G09,MIRES}.

\begin{figure*}[thbp]
\centering
\epsscale{1.1}
\plottwo{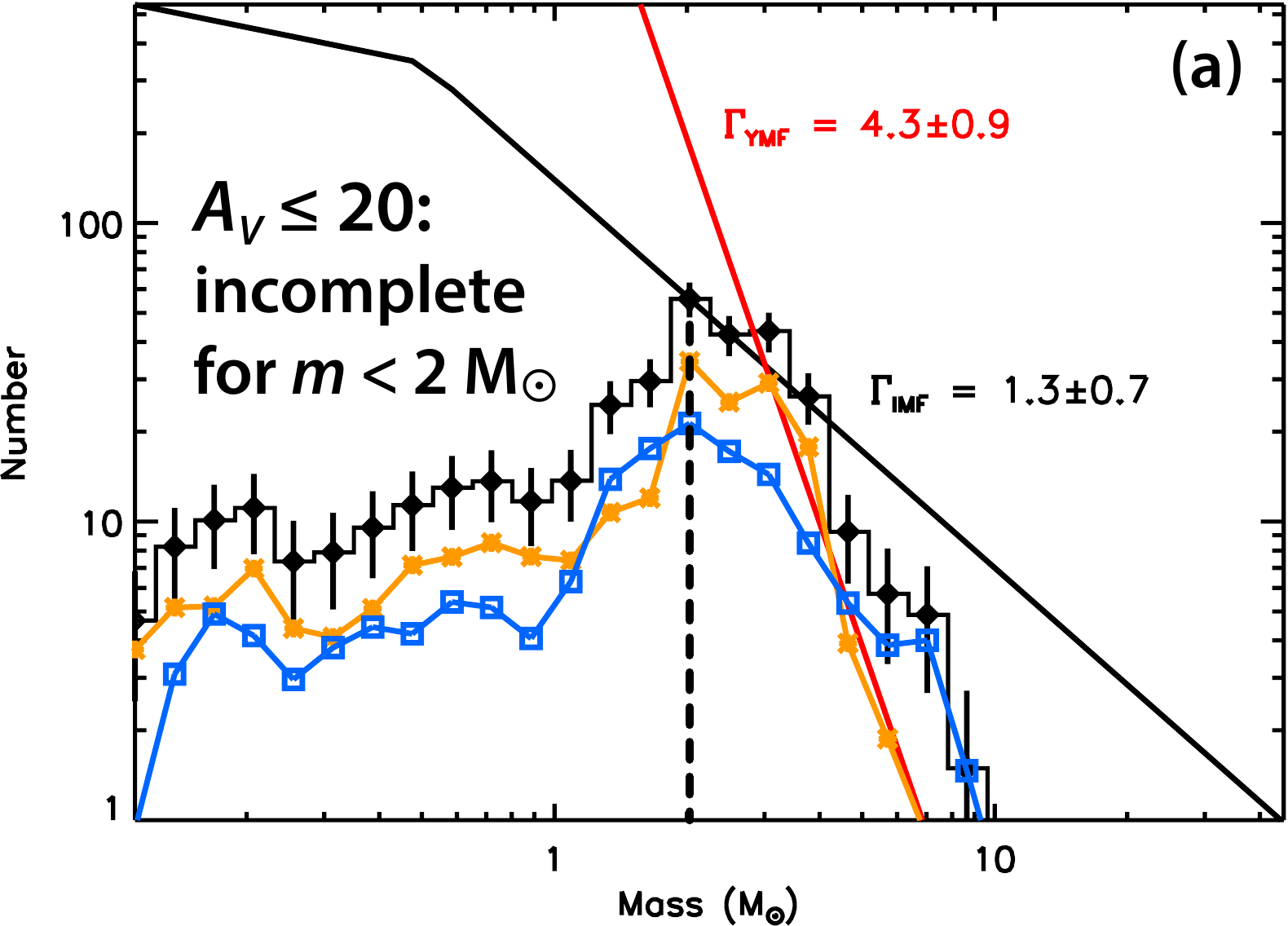}{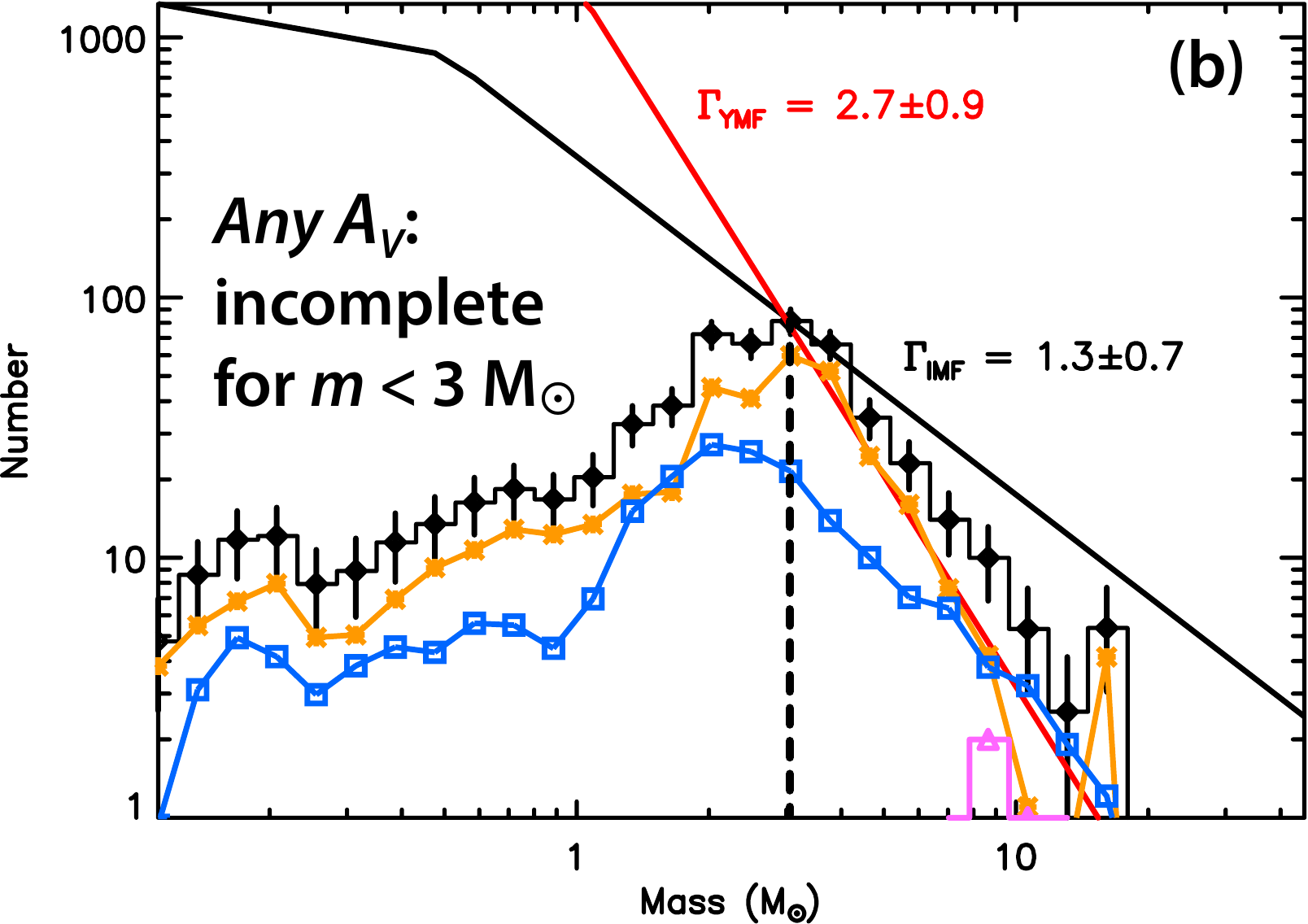}
\caption{Model mass functions plotted for (a) PCMs with $A_V^{\rm SED}\le 20$~mag and (b) all PCMs with SED fitting results. The orange curves show YSO mass functions (totaling 209 objects in panel a, 391 objects in panel b), exhibiting steepened power-law slopes at intermediate masses, as previously identified by PW10 (red line). The blue curves show the mass functions for X-ray PCMs with no IR excess (158 objects in panel a, 209 objects in panel b). 
The combined mass functions (black histograms with error bars) are used to extrapolate the mass of the total stellar population by scaling a Salpeter--Kroupa IMF (black curve). In panel (b) the total mass functions include masses estimated for the early B-types stars ionizing three compact MIR nebulae that were not included among our PCMs. The deficit of sources at low masses is due to photometric incompleteness, with approximate mass completeness limits indicated by dashed black lines and labelled on the figures. 
}
\label{fig:YMFs}
\end{figure*}

\subsection{Total Stellar Population}\label{sec:pop}

Following the approach used in numerous previous studies (e.g.\ \citealp{SP07}, \citealp{SAGEYSOs}, \citealp{P09}, PW10; \citealp{CCCPYSOs}) we use our SED fitting results to model  the YSO mass function (YMF) for the IR excess point-source population. In this work we introduce a new set of ``naked'' stellar models that we use to produce the analogous diskless mass function (DlMF) for the X-ray selected, non-IR excess population lacking inner dust disks (Appendix \ref{append:construct}). To probe the effects on our sample completeness of the high, variable extinction through the IRDC, in Figure~\ref{fig:YMFs} we again divide our PCMs by \avsed\ (see Figure~\ref{fig:avstrat}) and compare the mass functions of the less-obscured, $A_V^{\rm SED} < 20$~mag sample (panel a), to the mass functions of the {\em full} PCM sample for all \avsed\ (panel b).
In both panels DlMFs are plotted separately from the YMFs to highlight the striking differences in their shapes. 
The power-law slopes for the DlMFs are consistent with the standard Salpeter-Kroupa IMF, $dN/d\log{m}\propto m^{-\Gamma}$ with $\Gamma = \Gamma_{\rm IMF} = 1.3$ \citep{KW03}.
As PW10 found previously, a power-law fit to the intermediate-mass YMF returns a significantly steeper slope, $\Gamma_{\rm YMF} > 1.3$. 
For the less-obscured YMF in Figure~\ref{fig:YMFs}a, the best-fit $\Gamma_{\rm YMF}$ is so steep that we interpret it as a lack of {\em any} disk-bearing YSOs with $M_{\star} > 4$~\Msun\ and $A_V^{\rm SED}\le 20$~mag. The less-obscured YMF is consistent with the Salpeter-Kroupa IMF and the DlMF slope for $2~M_{\sun} < M_{\star} < 4~M_{\sun}$; both the YMF and the DlMF are incomplete at lower masses. The mass completeness limit of the YMF for the full sample (Figure~\ref{fig:YMFs}b) is higher, $M_{\star}\sim 3$~\Msun, revealing the impact of significant MIR extinction on our photometric sensitivity.

We sum the DlMFs and YMFs to produce the combined PCM mass functions, which contain 370 pre-MS stars and YSOs for the less-obscured sample and 603 in the full sample (black histograms in Figure~\ref{fig:YMFs}).
While the addition of diskless sources partially compensates for the steep YMF slope and brings the combined PCM mass function closer to the expected Salpeter-Kroupa slope, we do not expect completeness because
our sample is biased against X-ray quiet, late B stars that lack IR excess and have neither detectable intrinsic X-ray emission nor X-ray bright companions. As previously noted by PW10, the lack of massive, O-type stars is real, not a selection effect, and we will discuss its implications in Section~\ref{sec:massive}.

We follow our standard procedure for measuring the number of stars (above the hydrogen-burning limit at $M_{\star}\sim 0.1$~\Msun) in a model stellar population $N_{\rm pop}$ and their total mass $M_{\rm pop}$ (\citealp{SP07,SAGEYSOs,P09}; PW10; \citealp{CCCPYSOs}). We assume the actual mass function follows the Salpeter-Kroupa IMF scaled to match the peak of the observed mass function (Figure~\ref{fig:YMFs}). We then integrate over the scaled IMF, propagating uncertainties both from the scaling fit and on the IMF parameters. 
For the full PCM population in the ACIS FOV (Figure~\ref{fig:YMFs}b), we find $N_{\rm pop}=11,700\pm 2600$ and $M_{\rm pop}=7200\pm 1200$~\Msun. The actual population could be significantly larger, as our observed mass functions fail to sample completely the most highly-embedded YSOs in the IRDC, and miss other PCMs that happen to lie on the back side of the cloud, behind the IRDC filaments. For the combined mass function for the less-obscured population (Figure~\ref{fig:YMFs}a) we find $N_{\rm pop}(A_V^{\rm SED})=4700\pm 1000$ and $M_{\rm pop}(A_V^{\rm SED})=2900\pm 500$~\Msun. The less-obscured, less-clustered (Figure~\ref{fig:avstrat}) population thus represents ${\le}40\%$ of the full population traced by the observed PCMs.

\section{Discussion}\label{sec:discussion}

\subsection{Disk Destruction Timescales for Intermediate-Mass Pre-MS Stars}\label{sec:disks}
Our \chandra\ observation of M17 SWex has identified a substantial population of  
very young IMPS with no IR excess.
Clearly the inner disk lifetime for IMPS is shorter than the $\sim{3}$~Myr inner disk lifetime for low-mass T Tauri stars, as numerous previous studies have reported \citep{HerbigAeBe,HLL01,H07,W3diskfrac}. 
PW10 parameterized the disk lifetime $\tau_d(m)$ for IMPS in terms of a disk fraction with a power-law mass dependence, 
\begin{equation}\label{eq:diskfrac}
  \frac{\tau_d(M_{\star})}{\tau_{\rm SF}} = f_d(m) \propto M_{\star}^{-\delta};~~M_{\star}>1~M_{\sun},
\end{equation}
where $\tau_{\rm SF}$ is the time since star formation began in the complex, and $\delta = \Gamma_{\rm YMF} - \Gamma_{\rm IMF}$, the deviation of the observed YMF power-law slope from the IMF (Figure~\ref{fig:YMFs}). Note that the PW10 parameterization applies only where $\tau_d(M_{\star}) \le \tau_{\rm SF}$ and assumes a constant SFR durng that time interval, hence it differs fundamentally from the most widely-used definitions of disk lifetime in the literature, which generally assume that all stars formed simultaneously at time $\tau=0$, plot $f_d$ versus $\tau$ for various samples of stars at different (single) ages, and define $\tau_d$ as the age at which $f_d$ drops below a certain threshold value \citep[e.g.][]{HLL01,Y14}. Such assumptions are not valid for very young, massive complexes such as M17 SWex, which features a high rate of ongoing star formation, evident rapid disk evolution among IMPS, but a very high overall disk fraction (presumably $f_d$ approaches unity for the largely undetected, low mass T Tauri population). 

For the full M17 SWex YMF in Figure~\ref{fig:YMFs}b, $\delta = 1.4$. If we assume $\tau_d(1~M_{\sun})=3$~Myr \citep{HLL01}, Equation~\ref{eq:diskfrac} implies $\tau_d(4~M_{\sun})=0.4$~Myr and $\tau_d(10~M_{\sun})=0.1$~Myr. 
These values agree very well with reported Herbig Be disk lifetimes \citep{HerbigAeBe} and massive YSO ages \citep{RMS11b}, respectively.
Hence the PW10 power-law parameterization appears to hold for YSOs across the range from intermediate to high masses.


\subsection{Star Formation Rate}
The modeled diskless mass functions for the X-ray detected population peak at $M_{\star}\sim 2$~\Msun, before departing from a power-law due to incompleteness at lower masses (Figure~\ref{fig:YMFs}).
The power-law disk lifetime parameterization above (Equation \ref{eq:diskfrac}), predicts $\tau_{d}(2~M_{\sun})=1.1$~Myr, which provides an estimate for the duration of star formation, $\tau_{\rm SF}$, in M17 SWex.
The star formation rate (SFR) is thus $\dot{M}=M_{\rm pop}/\tau_{\rm SF}=0.0072$~\Msun~\peryr, and likely accelerating, given the larger size and younger age of the highly-obscured population. PW10 reported $\dot{M}=0.011$~\Msun~\peryr\ for the entire M17 SWex complex, but using a much shorter timescale of $\tau_{\rm SF} = 0.7$~Myr for the intermediate-mass YSO population only. Our value applies only to the central region, the G14.225--0.506 IRDC (the ACIS FOV), which contains ${\sim}50\%$ of the YSOs in our MIR excess source (MIRES) catalog (Figure~\ref{fig:ISED} and Table~\ref{tab:ised}). Extrapolating our results to all of M17 SWex would double the SFR, to $\dot{M}\approx 0.014$~\Msun~\peryr. Compared to the PW10 results, our incorporation of X-ray-selected, diskless stars increases both the size of the M17 SWex stellar population and the duration of star formation, with the net effect of an (insignificant) increase in the measured SFR.

\subsection{Distributed Star Formation Processes in Filamentary Molecular Clouds} 
While the spatial distribution of PCMs generally reflects the imprint of the IRDC filaments, the majority of detected PCMs are distributed outside of the several dense clusters, including the 40\% of PCMs in the less-obscured sample that are located outside the densest filaments (Figures~\ref{fig:members} and \ref{fig:avstrat}). The less-obscured population is likely older than the highly-obscured population, as it contains the majority of the diskless PCMs ($t_{\star} \le 2$~Myr) but no YSOs with $M_{\star}>4$~\Msun\ ($t_{\star}\ge \tau_d(4~M_{\sun})=0.4$~Myr). The vast majority of massive stars forming in M17 SWex, including all ${\sim}10$ massive stars and YSOs with $M_{\star}>10$~\Msun, are found in the highly-obscured population (see Figure~\ref{fig:YMFs}). 
Similar apparent age and mass gradients, in which low-mass YSOs are distributed along filaments and high-mass YSOs concentrated in dense cores, have been reported for the well-studied IRDC G34.4+00.24 \citep{SP07,F14}. Our incorporation of \chandra\ X-ray data and SED modeling now extends these trends to more-evolved IMPS lacking inner disks.

The observed ``filament-halo'' age gradient and mass segregation in M17 SWex could be explained by two interrelated star formation processes: 
\begin{enumerate}
\item {\em Filament-driven star formation with dynamical relaxation.} Stars form in the filaments, and the halo population is produced as older stars migrate outward via some rapid dynamical relaxation process(es). This is plausible given the short dynamical timescale in high-density filaments, and might even produce the observed mass segregation \citep{GH96,B+03}. 
\item
{\em Hierarchical filament collapse.} The less-massive halo population formed primarily from lower-mass clumps in the IRDC filaments and perhaps also in lower-density ``streamers'' in the halo gas flowing onto the IRDC filaments. 
The hierarchical collapse models of \citet{VS09} predict numerous, shallow gravitational potential wells surrounding a smaller number of deep potential wells. The shallower wells form lower-mass clusters early in the simulation, which subsequently can fall into the deep potential wells, increasing the gas mass collecting there and ultimately forming massive clusters.
As \citet{GFK14} point out, hierarchical collapse could explain observations that massive stars are among the last to form (\citealp{FT08}; \citealp{O10}), as they are born exclusively in the densest molecular cores.
\end{enumerate}

Simulations of star formation generally predict dynamical relaxation and evaporation caused by dynamical encounters between stars and expulsion of gas by protostellar jets, stellar winds, and expanding \hii~regions.  \citet{B+03} further predicted that dynamical interactions could truncate circumstellar disks and limit the mass of stars ejected from clouds by removing them from their supplies of accreting gas. Such a process could perhaps explain the filament-halo mass segregation observed in M17 SWex. The cloud simulated by \citet{B+03} was 2--3 orders of magnitude less massive than M17 SWex, and the ejected low-mass stars and brown dwarfs would be far below the detection limits of our observations, so it is not clear that their predictions of mass segregation can be extrapolated to the intermediate-mass stars in our PCM halo population. We note that ``diskless'' in our nomenclature refers to the lack of a MIR excess from an {\em inner} dust disk. 
It is possible (even likely) that our ``diskless'' stars still possess debris disks with large inner holes and excess emission at IR wavelengths longer than 24~\um. Our MIR data do not allow us to identify debris disks or to reliably constrain the outer radii of YSO disks, so we cannot make any statements about disk truncation. At present, hierarchical filament collapse models \citep{VS09,ZA+12,ZA+VS14} appear to provide a more natural explanation for the observed mass and age segregation in M17 SWex.

Compared to IRDC complexes dominated by a smaller number of long, linear filaments, for example the ``Nessie'' nebula \citep{Nessie} or G034.42+00.24 \citep{SP07} M17 SWex may be a relatively low-density GMC. M17 SWex may be similar to L1641, the southern region of the Orion A GMC, which is a very active star-forming region that has not produced massive stars \citep{M12,refereespaper14}.  Simulations by \citet{B11} predict an IMF weighted toward intermediate masses in lower-density clouds, in which case our low-mass IMF extrapolations using a standard Salpeter-Kroupa model would overestimate the true stellar mass and hence SFR in M17 SWex. Such a model is more likely to apply to the less-obscured population in the lower-density halo than to the high-density IRDC filaments themselves, suggesting that the extrapolated halo population could be overestimated. Our extrapolations predict that the (likely incomplete) younger, clustered population represents ${>}60\%$ of the {\em total} stellar population, hence we do not expect that our reported global SFR is significantly overestimated due to our assumption of a standard IMF.

\subsection{Prospects for Future Massive Star Formation in M17 SWex}\label{sec:massive}
As PW10 pointed out, M17 SWex seems poised to produce a large stellar association with a significant deficit of massive stars. The SFR measured within the ACIS FOV is $\dot{M}\ge 0.0036$~\Msun~\peryr, {\em at least} four times the SFR of the Orion Nebula Cluster (ONC), and the extrapolated SFR for the entire M17 SWex complex (recalling that only half of the YSOs in M17 SWex are concentrated in our ACIS FOV) rivals or exceed the $\dot{M}=0.005$~\Msun~\peryr\ SFR in NGC 6618, the cluster ionizing the bright M17 \hii~region itself \citep{CP11}. 
This is not an artifact of time-averaging, since the durations of star formation in M17 SWex, NGC 6618, and the ONC are approximately the same (\citealp{P09,CP11,GFK14}). 

The lack of O-type stars with $M_{\star}>20$~\Msun\ anywhere in M17 SWex hence remains difficult to explain in the context of a standard IMF (Figure~\ref{fig:YMFs}b). 
As PW10 noted, O stars inevitably heat the surrounding gas and dust, producing \hii regions that cannot be missed in MIR images, but we have either directly detected or included the equivalent masses of the ionizing stars responsible for all of the bright MIR nebulae within the ACIS FOV (see Figures~\ref{fig:Xcat} and \ref{fig:members}), none of which is more luminous than an early B star. 

The dearth of OB stars could be a statistical fluctuation, as the SFR of M17 SWex falls within the range where we might not expect complete sampling of the upper IMF \citep{KE12}. But based upon our modeling of the PCM population mass function (Figure~\ref{fig:YMFs}b) the OB deficit appears to be statistically significant. Using the lower-bound $N_{\rm pop}=9000$ derived from the mass function, a Salpeter-Kroupa IMF predicts between 11 and 50 OB stars with $M_{\star}\ge 20$~\Msun\ (the range is due to the reported uncertainty on $T_{\rm IMF}$; \citealp{KW03}) for the ACIS FOV alone, and extrapolating to all of M17 SWex predicts twice as many OB stars. 

The O-star deficit in M17 SWex seems extreme when compared to the local volume of the Galaxy within 500 pc of the Sun. The Orion molecular cloud, dominated by the ONC, contributes more than half of the SFR in the local volume and contains multiple O stars \citep{L10,KE12}. The SFR in M17 SWex is at least four times the SFR in the local volume, yet has not produced any O stars.

Because M17 SWex contains active sites of massive star formation, we should not discount the possibility that its massive cores may yet form O stars. None of the cores in M17 SWex appears completely quiescent (even the least active core, G014.077--00.558, harbors at least one highly-embedded protostar, visible in the MIPSGAL 24~\um\ image, that is not detected at shorter wavelengths, and a dense grouping of three YSOs and one X-ray source is found on its eastern edge; Figure~\ref{fig:members}).  The embedded star clusters ionizing compact \hii regions (IR-N, IR-S, and IR-SE) and MALT90 molecular cores containing YSOs and protostars seen only at 24~\um\ (see Table~\ref{tab:MALT90} and Figure~\ref{fig:members}) represent possible sites for massive cluster formation at a range of evolutionary development. 

\citet{B16} resolve Hub-N and Hub-S into 26 fragmentary 1.3 mm continuum sources, ranging in mass from ${<}1$~\Msun\ to 20~\Msun. None of these fragmentary sources {\em currently} is massive enough to form an O-type star. However, the total mass $M_{\rm tot}\sim 10^3$ of each hub is of the same order as
very massive cores observed in other IRDCs many of which harbor protoclusters with luminosities exceeding any of the protoclusters in M17 SWex \citep{D+11a,D+11b}.
The lower-density halo surrounding the IRDCs represents an enormous mass reservoir containing the bulk of the ${>2}\times 10^5$~\Msun\ measured by CO observations of M17 SWex \citep{ELD79}.
In other IRDCs, there is mounting evidence that lower-density halo gas continually falls inward onto IRDC filaments, flowing along the filaments and accreting onto the massive 
cores located at the filament nodes \citep{B14}. B13 note that Hub-N and Hub-S are locations where gas at different velocities converges. If such a process is underway in M17 SWex, the cores could rapidly gain mass, perhaps sufficient mass ultimately to form O stars in the protoclusters we already observe.

The hierarchical collapse models of \citet{VS09} lend theoretical support to the idea that massive complexes form their most massive stars relatively late in their star-forming history. As smaller clusters and cores fall into the larger cores, the most massive (proto)clusters can continue to grow, leading to an SFR that increases with time \citep{ZA+12,ZA+VS14}. Our result that the highly-obscured stellar population clustered along the cloud filaments is systematically younger and more massive than the more distributed halo population is fully consistent with an accelerating SFR.

\subsection{The Surprising X-ray Protostar Associated with EGO G14.33--0.64}\label{sec:surprising}
\begin{figure}[htp]
\epsscale{1.1}
\plotone{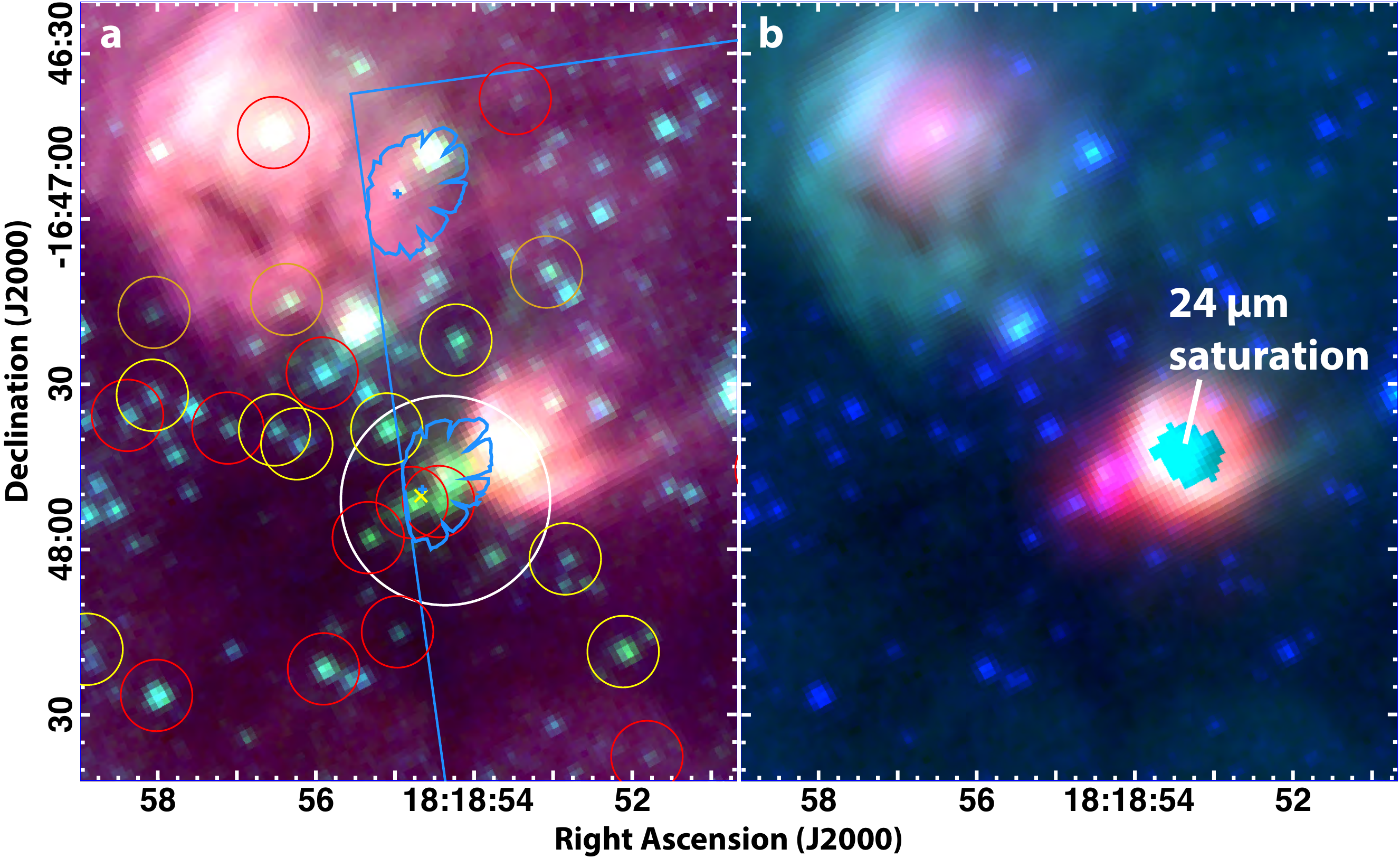}
\caption{Two MIR images showing the EGO G14.33--0.64 and environs: (a) red = 8.0~\um, green = 4.5~\um, blue = 3.6~\um\ and (b) red = 24~\um, green = 8.0~\um, blue = 4.5~\um. Overlays in panel (a) show positions of 2 X-ray sources (small blue 
``+'' symbols) with their photometry apertures (blue polygons), the border of our ACIS-I observation (blue lines), YSOs (circles color-coded by evolutionary stage as in Fig.~\ref{fig:ISED}), the MALT90 core (white circle) and the position of the \citet{S10} water maser (yellow ``${\times}$''). CXO J181854.64-164749.1 is the X-ray source closer to the center of the image.}
\label{fig:EGO}
\end{figure}

We have discovered a luminous, very hard X-ray source, CXO J181854.64-164749.1, at the extreme Northeast corner of our field (Figure~\ref{fig:EGO}). The source is in a massive protocluster associated with the UC \hii~region and \spitzer--identified extended green object G14.33--0.64 \citep[EGO;][]{egos}. 
Like CXO J181811.29-165722.2 discussed in Section~\ref{sec:spatial} above, CXO J181854.64-164749.1 appears to be associated with an extremely embedded, intermediate- or high-mass protostar, but it is even more obscured, ${\sim}3\times$ more luminous in X-rays, and associated with both a water maser and powerful, protostellar outflow \citep{C13}.

The position of CXO J181854.64-164749.1 coincides more closely with a water maser \citep{S10,C13} than with any MIR point source (Figure~\ref{fig:EGO}). Maser parallax measurements place G14.33--0.64 at $d=1.1$~kpc \citep{S10}, in strong disagreement with the kinematic distance to M17 SWex. Close inspection of high-resolution CO molecular line maps of the region (C. Beaumont, private communication) indicates that the G14.33--0.64 cluster is associated with a filament of gas that is kinematically distinct from the main body of M17 SWex, hence this appears to be a foreground cluster on the edge of our ACIS field. A nearer distance explains why the MIR emission from the G14.33--0.64 cluster is much brighter than that associated with any source in M17 SWex, while the luminosity of the cluster is actually similar to that of the various embedded clusters in the G014.225--00.506 IRDC. 
The MALT90 core mass derived from FIR dust emission is similar to the cores in G014.225--00.506 (Table~\ref{tab:MALT90}).

CXO J181854.64-164749.1 was detected with 28 net counts, all of them in the hard band (2--8 keV) with  $E_{\rm med}  =  5.11$~keV. Preliminary XSPEC fitting indicates $\log{N_H}=23.55$ and $\log{L_{t,c}}=31.7$ (assuming $d=1.1$~kpc), consistent with an intermediate- to high-mass star observed through $A_V>200$~mag of extinction. The source does not show evidence for variability. 
Because there is a relatively high uncertainty ($1\arcsec$) on the source position due to its extreme off-axis location, partially off the edge of the ACIS-I detector, a positive association of the X-ray source with a particular multiwavelength counterpart is not possible, but given its environs this source is most likely associated with an extremely young, intermediate-mass protostar.

\section{Conclusions}\label{sec:conclusions}

We have presented a detailed, multiwavelength study of the young stellar population associated with M17 SWex, built around a 100-ks \chandra\ observation of the IRDC G014.225--00.506. Our conclusions are summarized below.

\subsection{Science Discoveries}
\begin{enumerate}
  \item Rapid disk evolution on ${<}1$~Myr timescales has produced a significant population of IMPS lacking inner dust disks, as predicted by P10 and now revealed by their detected X-ray emission and lack of IR excess emission. This supports previous results that the disks around the intermediate-mass progenitors to Herbig Ae/Be stars are destroyed more quickly than those of their low-mass T Tauri analogs \citep{HerbigAeBe,HLL01}. We find that the disk lifetime can be parameterized as a negative power-law function of stellar mass with an index $\delta = 1.5$.
  \item We observe ``filament--halo'' gradients in both mass and age of the stellar population. A heavily-obscured population tightly clustered along the IRDC filaments contains all of the youngest and most massive YSOs. A more distributed, less-obscured population in the lower-density molecular halo contains the majority of the diskless IMPS.
  \item Based on the presence of 2--3~\Msun~IMPS without inner dust disks, M17 SWex has been an active SF region for the past ${\ga}1$~Myr, longer than might have been expected for an IRDC, which are generally regarded as providing the observational laboratories for the initial conditions for massive star formation. 
  \item Our \chandra\ observation has detected three hard, luminous X-ray sources associated with highly-embedded intermediate- or high-mass stars: CXO J181825.47-165618.4, CXO J181811.29-165722.2, and CXO J181854.64-164749.1. These appear to be among the youngest stellar X-ray sources known. The most luminous of this group is CXO J181854.64-164749.1, discussed in Section~\ref{sec:surprising}. This source is associated with protostellar outflow activity, traced by a MIR ``EGO'' source \citep{egos} and a water maser \citep{S10} in a foreground cluster on the corner of our ACIS FOV. We will present a detailed study of the X-ray spectra of these and other remarkable sources associated with high-mass stars, protostars, or IMPS in a subsequent paper.
\end{enumerate}

\subsection{Open Questions}
\begin{enumerate}
\item What produces the observed ``filament--halo'' age gradient? Our leading scenario is {\em filament-driven star formation with dynamical relaxation} coupled with {\em hierarchical filament collapse}, but the relative importance of each process in producing the observed spatial and temporal distributions of stars is not clear. 
\item Are massive stars born late? M17 SWex currently hosts no O-type stars, a significant deficit of massive stars compared to the predictions of a standard IMF, but the cloud provides a large mass reservoir that could eventually produce massive clusters.
\end{enumerate}

Our two open questions are interrelated.
Hierarchical collapse models \citep{VS09} predict that massive stars form last, ultimately populating the high-mass IMF once massive cores have had time to accrete sufficient mass from the surrounding filaments and halo gas \citep{B14}.
Ongoing filament collapse is the only scenario that would enable the massive cores in M17 SWex to accrete ${\sim}5$--10 times their current mass and form early O stars. If this can happen, then \citet{EL77} were right in their original prediction that M17 SWex represents the next generation of massive star formation in the M17 complex.

The gas dynamics in M17 SWex could answer the above questions. This would require more detailed kinematic studies that simultaneously employ both low-density tracers like CO and high-density tracers like NH$_3$. But given the complex morphology of M17 SWex, with its multiple, overlapping filaments, the gas kinematics may be difficult to interpret. Deep X-ray observations could reveal similar evolved, diskless stellar populations and hence age gradients in other, more well-studied IRDCs with simpler morphologies. This could provide easier observational tests of the relationship between stellar populations and gas dynamics.

Even if hierarchical filament collapse drives star formation in M17 SWex, the massive cores may fail to accrete sufficient mass to form truly massive, O-star clusters. M17 SWex could be a prototype for a larger class of extended, massive, yet relatively low-density star-forming clouds, characterized by high SFRs but a paucity of dense clusters ionizing bright \hii regions.  Another potential member of this class is L1641, the famous ``hockey stick''-shaped southern extension of the Orion A GMC, which contains more than 2000 {\em Spitzer}-detected YSOs but no O-type stars  \citep{M12,refereespaper14}.  If such clouds are common, they could represent a significant mode of ``dark'' star formation, invisible to widely-used SFR diagnostics in external galaxies \citep{CP11,KE12}.

\acknowledgements We are grateful to J. B. Foster for providing data and analysis for the MALT90 cores in M17~SWex and to C. Beaumont for sharing the JCMT CO $(J=3\rightarrow 2)$ datacube. We thank L. A. Hillenbrand for suggesting the $A_V$ stratification of the PCM sample, C. Battersby for helpful discussions on the implications of the missing OB stars in M17 SWex, and E. V\'{a}zquez-Semadeni for the insight that either delayed massive star formation or low-density star-forming clouds are consistent with existing simulations. We thank the anonymous referee for numerous constructive suggestions that improved the clarity and presentation of this work. M.S.P. was supported by an NSF Astronomy and Astrophysics Postdoctoral Fellowship under award AST-0901646 during the principal analysis phase of this project and is currently supported by the NSF through grant CAREER-1454333. This work was supported by {\em Chandra X-ray Observatory} general observer grant GO1-12016X (Science PI: M. Povich). This work is based in part on
observations from the {\it Spitzer Space Telescope} which is operated by the Jet Propulsion Laboratory, California Institute of Technology under a contract with NASA.
This work is based in part on data obtained from the United Kingdom Infrared Telescope (UKIRT) as part of the UKIRT Infrared Deep Sky Survey, when UKIRT was operated by the Joint Astronomy Centre on behalf of the Science and Technology Facilities Council of the UK.
This publication makes use of data products from the Two Micron
All-Sky Survey, which is a joint project of the University of
Massachusetts and the Infrared Processing and Analysis
Center/California Institute of Technology, funded by NASA and the NSF. 

{\em Facilities:} \facility{CXO (ACIS)}, \facility{Spitzer (IRAC)},  \facility{Spitzer (MIPS)},  \facility{CTIO:2MASS}, \facility{UKIRT (WFCAM)}

\appendix

\section{Identifying Probable Complex Members}\label{append:PCMs}

In this Appendix we summarize our procedures for cross-matching the various IR and X-ray catalogs and subsequently analyzing the properties of IR and X-ray sources to identify and characterize YSOs and diskless PCMs detected in X-rays.

\subsection{MIR Identification of Young Stellar Objects Associated with M17 SWex}\label{IRE}

Updating the work of PW10, we use the methodology of \citet[][hereafter P13]{MIRES} to identify infrared excess sources from the broadband 1--8~\um\ photometry and define a subsample of candidate YSOs that are probable members of the M17 SWex complex. Table~\ref{numbers} summarizes the numbers of {\em Spitzer} MIR sources involved in each step of this selection process for the M17 SWex wide field. Additionally, for candidate YSOs included in this MIRES catalog (Table~\ref{tab:ised}) we report number counts of YSOs contained within the smaller ACIS-I field-of-view and (XFOV = 1) and which of those were positionally matched to an X-ray source (X\_det = 1).


\begin{deluxetable}{lrrrl}
\tabletypesize{\scriptsize}
\tablecaption{Summary Counts of GLIMPSE Point Sources Used for YSO Identification\label{numbers}}
\tablewidth{0pt}
\tablehead{
  \colhead{Sources} & \colhead{Wide Field} & \colhead{ACIS FOV} & \colhead{X-ray Matched} & \colhead{Description}
}
\startdata
In GLIMPSE Catalog      & 150,523 & & & Detected IRAC point sources, 99.5\% reliable \\
Fit with SED Models     & 141,889 & & & Detected in ${\ge}4$ of 7 $JHK_s$+IRAC bands \\
Stellar Photospheres    & 115,264 & & & Well-fit by stellar atmosphere SEDs \\
Possible IR Excess      & 26,625 & & & Poorly-fit by stellar atmosphere SEDs \\
Marginal IR Excess      & 24,519 & & & Apparent excess in only IRAC [5.8] or [8.0] band \\
Significant IR Excess    &  2106 & & & Non-stellar; MIPS 24~\um\ photometry performed \\
\hline   
Candidate YSOs          & 1207 & 352 & 101 (29\%)  & MIRES catalog, corrected for contamination \\
PCMs                    &  698 & 334 & 98 (29\%)  & YSO probable complex members \\
~~~~-- Stage 0/I               &  253 & 115 & 23 (20\%)  & Dominant SED models: infalling envelopes \\
~~~~-- Stage II                &  353 & 171 & 63 (37\%)  & Dominant SED models: circumstellar disks only \\
~~~~-- Ambiguous               &  100 & 48  & 12 (25\%) & No dominant SED model type
\enddata
\end{deluxetable}

As a preliminary step, we performed a 1\arcsec\ radius positional matching of 667,325 high-quality\footnote{See \citet{NIRoptimal} and P13 for details.} UKIDSS sources to 150,523 GLIMPSE Catalog sources in our IR Wide Field. If multiple UKIDSS sources fell within the matching radius of a given GLIMPSE Catalog source, the closest UKIDSS source was adopted as the ``primary'' match, and the number of ``secondary'' possible matching sources was recorded. 
139068 (92\%) of the GLIMPSE Catalog sources had primary matches to UKIDSS, of which only
432 (${\sim}0.3\%$) were accompanied by secondary matches. For our subsequent photometric analyses, we preferred UKIDSS photometry wherever available, but substituted {\em 2MASS} photometry if necessary, mainly in cases of bright stars that saturated the UKIDSS images or sources flagged as affected by artifacts.

The remaining steps were adjusted from P13 to include MIPS 24~\um\ photometry:
\begin{enumerate}
\item Using the SED fitting tool of \citet{fitter}, we fit \citet{Kurucz} stellar atmosphere models, incorporating interstellar reddening as a free parameter ($0 \le A_V^{\rm SED} \le 45$~mag; the upper limit set by inspection of the near-IR colors of all point sources detected in the field, see P13) using the extinction law of \citet{I05}, to all sources with $N_{\rm data}\ge 4$ detections in the 7 combined NIR--MIR bands. Following P13, we set the {\em minimum} photometric error bars used in SED fitting to 5\% in the $JHK_S$, [3.6], and [4.5] bands and 10\% for [5.8] and [8.0].  Stellar atmosphere models provided poor fits ($\chi^2_0/N_{\rm data} > 2$; $\chi^2_0$ is the goodness-of-fit parameter for the best-fitting model) to 26,625 sources; this is more than an order of magnitude more than the 1498 sources found by PW10 at the same step. PW10 used only {\em 2MASS} NIR photometry; our inclusion of the deeper UKIDSS photometry allowed us to fit models to more than twice as many sources. PW10 also used a significantly more liberal cutoff for well-fit stellar atmospheres ($\chi^2_0/N_{\rm data} > 4$), and this difference is the primary reason that many more sources pass through our effectively coarser stellar-atmosphere filter (since sources that are well-fit by stellar atmospheres are removed from consideration as YSOs). This potentially allows us to identify more YSOs with weaker MIR excesses compared to PW10.

\item The large majority (24,519) of sources that were poorly-fit by stellar atmospheres are ``marginal IR excess'' sources that we remove using a set of color cuts 
designed to identify sources with apparent MIR excess in only a single band, usually the [5.8] or [8.0] band \citep{SP10,MIRES}. This source population is dominated by contaminating field stars and excluded from further analysis, with the following exception: marginal IR excess sources in the GLIMPSE Archive that are {\em counterparts to X-ray sources} are used for X-ray source classification (see Appendix~\ref{sec:classify}).

\item Following PW10 and \citet{CCCPYSOs}, we perform aperture photometry on the MIPSGAL 24~\um\ mosaics at the positions of the 2106 GLIMPSE Catalog sources that passed through the preceding two filtering steps. We detected and extracted 777 sources at 24~\um, placed upper limits on the 24~\um\ flux density for 1314 other sources, and found 15 sources that were saturated at 24~\um, for which we do not report or use 24~\um\ photometry.\footnote{Bright, massive YSO candidates in M17 SWex detected by {\em The Midcourse Space Experiment} (\msx; \citealp{MSX}) were previously cataloged by PW10; we incorporate three of these sources into our analysis of the M17 SWex stellar population in the ACIS FOV (Section~\ref{sec:PCM}).} We set the minimum [24] photometric errors to 15\% for SED fitting.

\item We identify 124 candidate 4.5~\um\ excess sources consistent with unresolved, shocked molecular line emission \citep[e.g.][]{egos} using the color selection criteria
$[3.6] - [4.5] > 1.1$ and $[3.6] - [4.5] > (1.9/1.2)\times ([4.5] - [5.8])$ (P13). For these sources the [4.5] flux density was used as an upper limit in subsequent SED fitting.

\item We fit the 1--24~\um\ SEDs of the 2106 remaining sources with \citet[][hereafter RW06]{grid} YSO models, assuming a distance of $d=2.0$~kpc and allowing for interstellar extinction $A_{V}<100$~mag\footnote{This essentially leaves $A_V$ unconstrained when fitting sources with YSO models, necessary because candidate protostars are found deeply embedded in the IRDC lanes, where extinction can be extreme and background stars are obscured even in the IR.}; 1203 sources were well-fit ($\chi^2_0/N_{\rm data}\le 4$). We visually reviewed the SEDs and locations in our IR images of the 903 sources that were poorly-fit by YSO models, and judged that four of them were legitimate candidate YSOs (see P13 for details of our visual review criteria). Hence we arrive at our final catalog containing 1207 MIRES in the M17 SWex Wide Field (top panel, Figure~\ref{fig:ISED}), analogous to the P13 catalog of ${\sim}20,000$ MIRES identified in 18 other Galactic massive star-forming regions (including M17 itself). The MIRES catalog is published as an electronic table accompanying this article, and its format is presented in Table~\ref{tab:ised}.

\item The most important difference between the format of our new MIRES catalog for M17 SWex and the MYStIX-based MIRES catalog presented by P13 is the inclusion of 24~\um\ photometry in Columns 4 \& 5 of Table~\ref{tab:ised}, which in turn enables us to define a new flag (SED\_flg = 4; Column 8) identifying candidate dusty, asymptotic giant branch (AGB) stars with $[8.0]-[24] < 2.2 $~mag \citep{SAGEYSOs,P09,CCCPYSOs}. Fifty candidate AGB stars (marked by cyan $\times$ symbols in Figure~\ref{fig:ISED}) were hence removed from consideration as YSO members of M17 SWex. The other values for SED\_flg were defined and used exactly as in P13, but are not very important to our analysis, since only 23 sources are flagged as starburst galaxies (the majority of which become candidate YSOs in the next step), and none as obscured AGN. 

\item Finally, we identify probable YSO members of M17 SWex (Column 13 of Table~\ref{tab:ised}) using the spatial distribution analysis described in detail by P13. The Prob\_dens parameter is computed for all MIRES within the defined M17 SWex Target Area (blue boundary in Figure~\ref{fig:ISED}) by scaling the surface density of MIRES in an off-cloud ``control'' field (outside the Target Area) relative to the surface density of non-IR excess sources throughout the entire Wide Field. Among MIRES initially flagged as YSOs (SED\_flg = 0), 698 with Prob\_dens$\ge 0.68$ are selected as probable members. An additional 8 MIRES initially flagged as galaxies (SED\_flg = 1) and with Prob\_dens$\ge 0.72$ are also selected as probable members, and hence should be regarded as candidate YSOs (hence in Table~\ref{tab:ised}, Column 13 overrides Column 8 for source classification).

\end{enumerate}

\begin{deluxetable}{rlp{4.75in}}
\tabletypesize{\small}
\tablewidth{0pt}
\tablecaption{ \label{tab:ised}
Mid-IR Excess Source (MIRES) Catalog Format}
\tablehead{ & Column Label & Description}
\startdata
(1) & MIR\_Name &  Source name in GLIMPSE Catalog \\  
(2) & RAdeg     & Right ascension (J2000, degrees) \\
(3) & Dedeg     & Declination (J2000, degrees) \\
(4) & IRmag     & Magnitudes in 8 IR bands used for SED fitting: $J$, $H$, $K_S$, [3.6], [4.5], [5.8], [8.0], [24] \\
(5) & IRmag\_err & Uncertainties on the 8 IR magnitudes used for SED fitting,\tablenotemark{a} {\em reset to floor values}\tablenotemark{b} \\
(6) & NIRphot\_cat   & Provenance of {\em each} of 3 $JHK_S$ sources matched to IRAC source: 0={\it 2MASS}, 1=UKIRT, $-1$=none \\
(7) & NIRphot\_num\_SM   & Number of UKIRT sources providing possible alternative, ``secondary'' matches to the IRAC source \\
(8) & SED\_flg & Source classification flag: 0=likely YSO, 1=starburst galaxy, 2=AGN, 4=candidate dusty AGB star \\
(9) & SED\_chisq\_norm & $\chi^{2}_{0}/N_{\rm data}$ of best-fit SED model.\tablenotemark{c} \\
(10) & SED\_AV & Visual extinction $A_V^{\rm SED}$ determined from $\chi^2$--weighted mean of all acceptable SED fits \\
(11) & SED\_stage & Evolutionary Stage classification, RW06 YSO models: 1=Stage 0/I, 2=Stage II/III, $-1$=ambiguous\tablenotemark{d} \\
(12) & Prob\_dens & $=1-f_{\rm con}$, where $f_{\rm con}$ is the fraction of MIRES in the local neighborhood that are consistent with foreground/background contaminants.\tablenotemark{e} \\
(13) & MEM\_flg & =1 if probable member of M17 SWex, 0 otherwise \\
(14) & XFOV & =1 if source falls within the {\em Chandra}/ACIS-I field-of-view, 0 otherwise \\
(15) & X\_det & =1 if MIR source is matched to an X-ray source, 0 otherwise \\
\enddata
\tablenotetext{a}{Value of $-99.99$ means the corresponding flux measurement was used as an upper limit for SED fitting.}
\tablenotetext{b}{As described in Section \ref{IRE}, the uncertainties on flux densities $F_i$ used for SED fitting were $\delta F_i \ge 0.05 F_i$ for $JHK_S$, [3.6], and [4.5], $\delta F_i \ge 0.10 F_i$ for [5.8] and [8.0], and $\delta F_{24} \ge 0.15 F_{24}$ for [24]. 
}
\tablenotetext{c}{For SEDs exhibiting shock-produced 4.5~\um\ excess emission, the [4.5] flux was used as an upper limit and $3\le N_{\rm data}\le 6$. For all other SEDs $4 \le N_{\rm data}\le 7$.}
\tablenotetext{d}{All MIRES catalog sources, regardless of SED\_flg, were fit with RW06 models and hence can be classified according to YSO evolutionary Stage.}
\tablenotetext{e}{``NaN'' values are assigned to MIRES catalog sources falling within designated ``control'' fields for clustering analysis.}
\end{deluxetable}

Thanks to the inclusion of deeper, UKIDSS NIR photometry and our updated methodology, we classified 706 YSOs as probable members of the M17 SWex IRDC complex (shown in the bottom panel of Figure~\ref{fig:ISED}), a 45\% expansion of the PW10 sample. Just under half (350) of these candidate YSOs are detected and unsaturated at 24~\um. As summarized in Table~\ref{numbers}, using the most probable RW06 YSO evolutionary stage classifications (SED\_stage, as defined by P13), the MIRES--M17SWex members are comprised of 253 (36\%) candidate protostars (Stage 0/I), 353 (50\%) disk-dominated YSOs (Stage II), and 100 (14\%) YSOs of ambiguous stage (poorly-constrained SED models, frequently due to lack of a 24~\um\ detection). 

\subsection{IR Counterparts to X-ray Sources}\label{sec:counterparts}
Our identification of IR counterparts and subsequent classification analysis of the X-ray source population in M17 SWex follows the strategies developed for the {\em Chandra} Carina Complex Project \citep[CCCP,][]{CCCP} and Massive Young Star-Forming Complex Study in Infrared and X-rays \citep[MYStIX,][]{MYStIX}. The principal innovation presented in this study is the deployment of a new grid of $10^5$ SED models for ``naked'' pre-MS and MS stars lacking dusty, circumstellar disks, which parallel the RW06 YSO models (see Appendix~\ref{append:construct} for details). We can hence use the \citet{fitter} SED fitting tool to model the properties of X-ray-selected members that do not exhibit strong IR excess from circumstellar disks.

Spatial matching between X-ray source catalog and IR counterpart catalogs used the procedure employed by \citet{CCCPCat} for the CCCP. The matching algorithm, described by \citet{AE}, calculates the maximum acceptable angular separation between an X-ray source and a counterpart that would generate ${<}0.1\%$ chance of a match due to a chance alignment. The algorithm uses the individual source positional uncertainties to calculate the probability of chance alignments (important because the \chandra\ PSF varies across the ACIS FOV). X-ray sources were matched separately to our custom UKIDSS optimal-photometry catalog and the GLIMPSE Archive list (the latter contains 2MASS photometry, where available, matching each IRAC source). UKIDSS NIR counterparts were identified for 582 (69\%) of X-ray sources, with a median separation of 0.36\arcsec.
GLIMPSE MIR counterparts were identified for 461 (55\%) of X-ray sources, with a median separation of 0.37\arcsec.

For SED fitting of IR counterparts to X-ray sources, if a high-quality, unsaturated UKIDSS counterpart was available it supplied $JHK_s$ photometry, otherwise 2MASS photometry was substituted if available via the GLIMPSE Archive. We ran these sources through the classification procedure described in Section \ref{IRE} and report the results in an electronic table accompanying this article, described in Table~\ref{tab:xsed}. 
MIPS 24~\um\ aperture photometry was performed for all X-ray-detected MIRES, as described in Appendix~\ref{IRE}.
Among our 152 X-ray detected YSOs, 98 were independently selected for our MIRES catalog (flagged in column 15 of Table~\ref{tab:ised}), and 54 were new identifications using the more complete GLIMPSE Archive photometry.
The evolutionary stages of the X-ray detected YSO population break down as 34 (22\%) Stage 0/I, 95 (62\%) Stage II/III, and 23 (15\%) with ambiguous stage (Table~\ref{tab:xsed}, column 14).

\begin{deluxetable}{rlp{4.75in}}
\tabletypesize{\small}
\tablewidth{0pt}
\tablecaption{ \label{tab:xsed}
Basic X-ray and IR Counterpart Properties Used in Source Classification
}
\tablehead{ & Column Label & Description}
\startdata
(1) & Xray\_Name &  ``CXO J'' X-ray source name \\ 
(2) & RAdeg     & Right ascension of X-ray source (J2000, degrees) \\
(3) & Dedeg     & Declination of X-ray source (J2000, degrees) \\
(4) & MIR\_Name & Source name in GLIMPSE Archive\tablenotemark{a} \\ 
(5) & UKIRT\_label   & UKIRT catalog source matched to X-ray source \\
(6) & IRmag     & Magnitudes in 8 IR bands used for SED fitting: $J$, $H$, $K_S$, [3.6], [4.5], [5.8], [8.0], [24] \\
(7) & IRmag\_err & Uncertainties on the 8 IR magnitudes used for SED fitting,\tablenotemark{b} {\em reset to floor values}\tablenotemark{c} \\
(8) & NIRphot\_cat   & Provenance of near-IR source matched to IRAC source: 0={\it 2MASS}, 1=UKIRT, $-1$=none \\
(9) & NIR\_excess  & =1 if NIR colors are consistent with $K_S$ excess emission, see Figure~\ref{fig:JHK}. \\
(10) & SED\_model\_type & Type of SED model fit to source: 0=naked (pre-)MS stellar models, 1=RW06 YSO models, $-99$=none \\
(11) & SED\_flg & Source classification flag: $-2$=stellar photosphere, $-1$=marginal IR excess, 0=likely YSO, 1=starburst galaxy, 2=AGN, $-99$=no acceptable SED fits \\
(12) & SED\_chisq\_norm & $\chi^2/N_{\rm data}$ of best-fit SED model, number of data points fit is $3\le N_{\rm data}\le 7$ \\ 
(13) & SED\_AV & Visual extinction \avsed\ determined from $\chi^2$--weighted mean of all acceptable SED fits \\
(14) & SED\_stage & Evolutionary Stage classification, RW06 YSO models: 1=Stage 0/I, 2=Stage II/III, $-1$=ambiguous, $-99$=unclassified\tablenotemark{d} \\
(15) & MedianEnergy & Total-band X-ray median energy \\
(16) & Var\_flg & =1 if ``definite'' X-ray variability detected\tablenotemark{e} \\
(17) & Class & Classification flag of X-ray source: 1 = foreground, 2 = PCM, 3 = background, 0 = unclassified
\enddata
\tablenotetext{a}{MIR\_Name is used for cross-indexing with MIRES; there are 101 X-ray sources with MIRES matches (identified in column 16 of Table~\ref{tab:ised}).}
\tablenotetext{b}{As in Table~\ref{tab:ised}, a value of $-99.99$ means that flux was used as an upper limit for SED fitting.}
\tablenotetext{c}{See Appendix~\ref{IRE} and Table~\ref{tab:ised}}
\tablenotetext{d}{Sources with SED\_flg $< 0$ were not fit with RW06 models and hence cannot be classified according to YSO evolutionary stage.}
\tablenotetext{e}{See Appendix~\ref{sec:variability} for the definition of ``definite X-ray variability.''}
\end{deluxetable}

We incorporate the P13 classifications of ``stellar photosphere'', ``marginal IR excess'', and ``no acceptable SED fits'' for non-MIRES sources (Table~\ref{tab:xsed}, column 11). We fit the SEDs of non- and marginal-IR excess sources (excluding any available 5.8 and 8.0 \um\ photometry from the fitting for the latter, because it is assumed to be unreliable) with our naked stellar models (see Appendix~\ref{append:construct}), setting $d=2.0$~kpc and allowing $A_V\le 50$~mag. The choice of SED model (YSO, naked, or none) grid used for each X-ray counterpart is reported in column 10. The SED fitting results using the appropriate model grid were used to compute \avsed, the $\chi^2$-weighted mean interstellar extinction ($A_{V}$), as defined by P13, to each source (column 13).

The 706 candidate YSOs that survived the spatial clustering cut against contaminating sources (Figure~\ref{fig:ISED}, bottom panel) are considered probable members of the M17 SWex association, and 334 of these are located within the ACIS field (MEM\_flg = 1, XFOV = 1, columns 13 and 14  of Table~\ref{tab:ised}), with one corner removed to exclude the G14.33-0.64 probable foreground cluster.

To identify probable complex members (PCMs) among the X-ray sources, we developed a simplified version of the membership classification scheme of \citet{CCCPClass,MYStIXPCM}. Our empirical approach substitutes binary decisions in place of Bayesian likelihoods based on various observable source properties, and it removes the need for simulating the expected X-ray source population in the field \citep[e.g.][]{CCCPSim}, since producing an accurate simulation would be very difficult given the spatially complex pattern of highly variable extinction produced by the IRDC. We also expect that the high absorption produced by M17 SWex itself and the long sightline through the Galactic mid-plane behind it will effectively screen out the majority of background contaminants (particularly AGN). We will evaluate quantitatively the impact of this expectation on our results using the observed spatial distribution of the X-ray sources with different source classifications (Section~\ref{sec:PCM}).

\subsection{Criteria for X-ray Source Classification}\label{sec:classify}
We classify X-ray sources as PCM or foreground stars based on the criteria described below. In Table~\ref{Xclass} we list, for each step in the classification procedure, the number of {\em new} classifications made along with the number of sources remaining unclassified.

\subsubsection{IR excess emission}
\begin{figure}[thp]
\epsscale{0.49}
\plotone{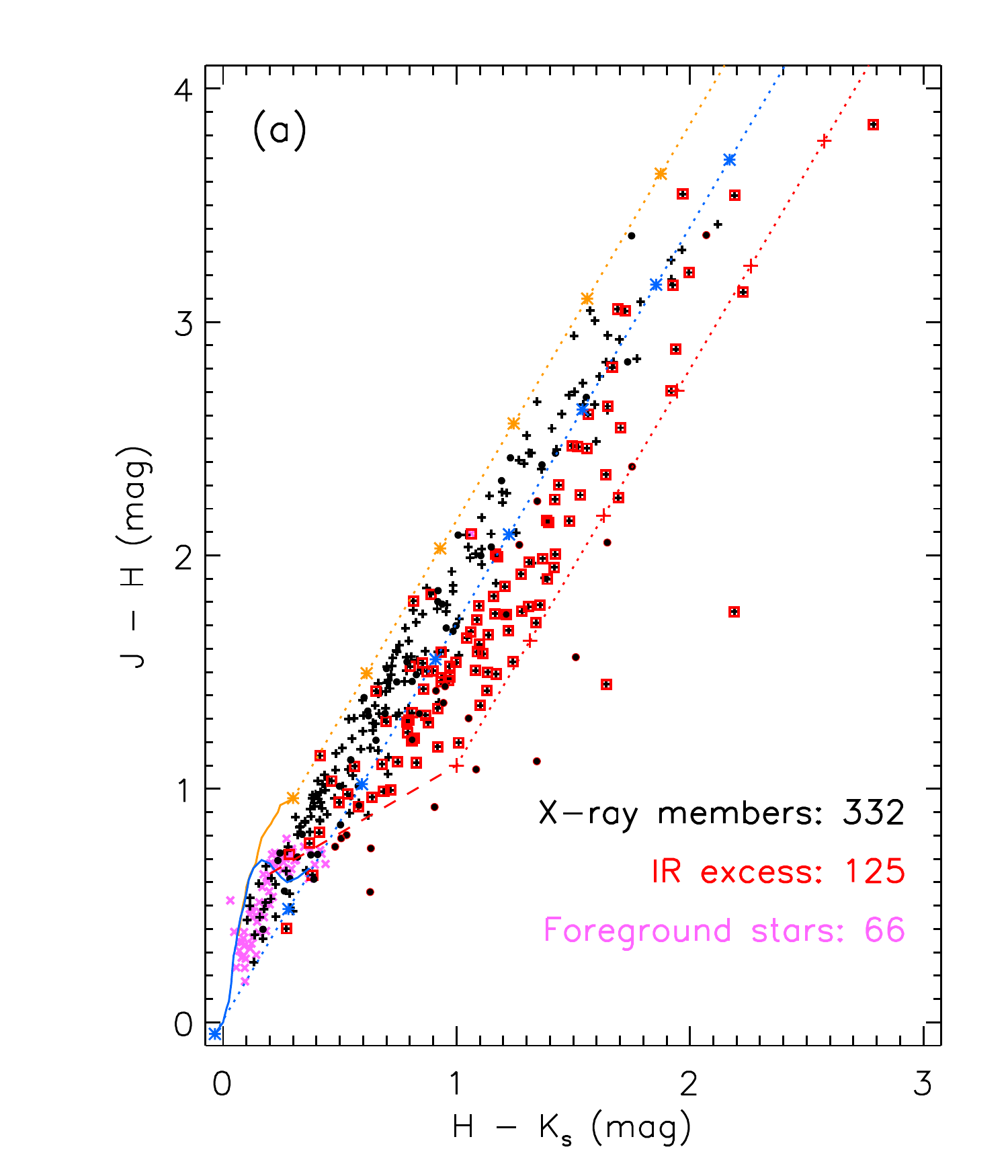}
\plotone{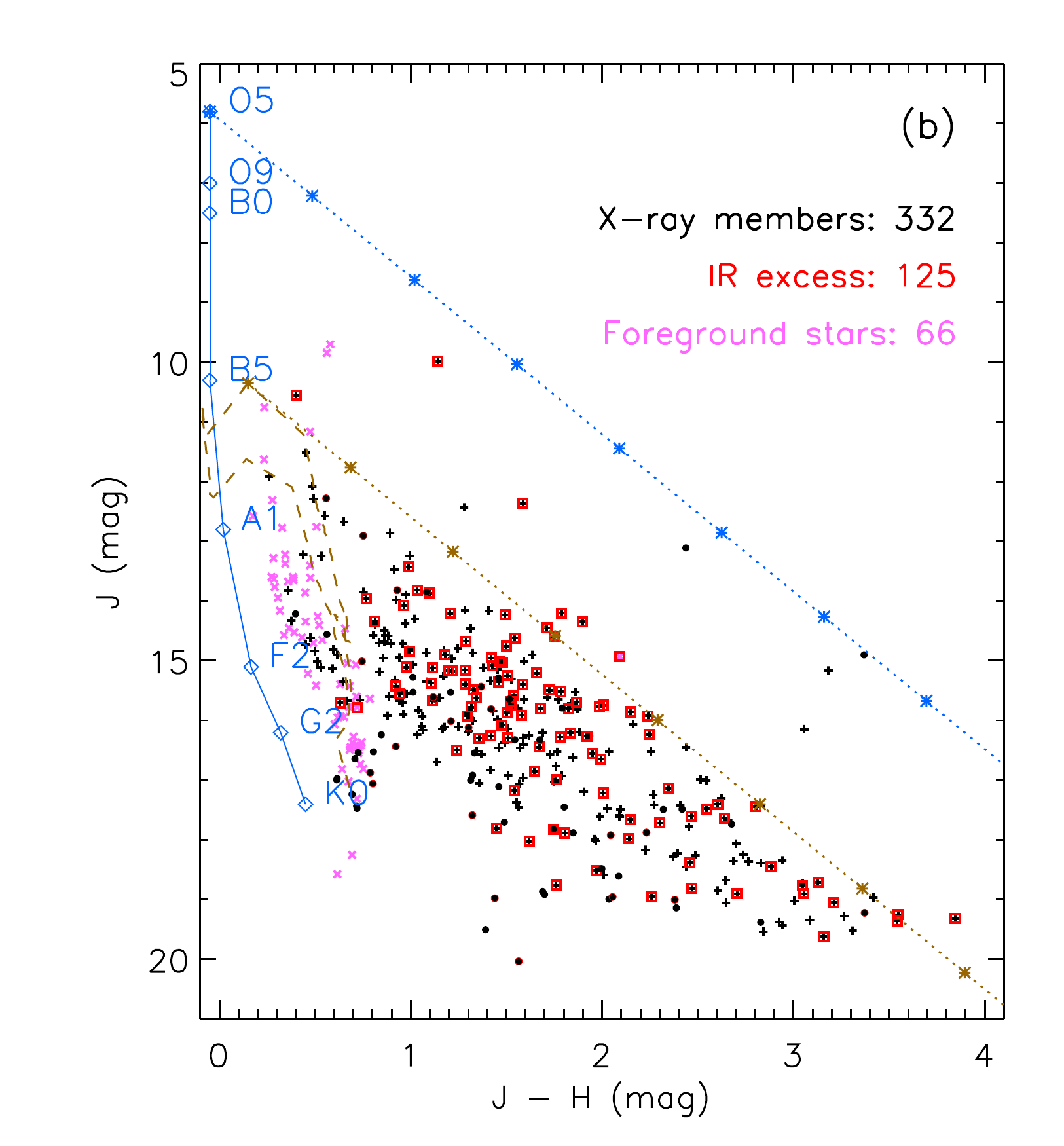}
\caption{({\it a}) $JHK_S$ color-color diagram of NIR counterparts to X-ray sources. Only sources with $J$ and $H$ photometric uncertainties ${\le}0.1$ mag are plotted. Sources with SED fitting results are plotted black crosses, with red squares added for YSOs with MIR excess emission. The remaining NIR counterparts that lacked SED fitting results are plotted as black dots, with red circles added for sources with $K_S$ excess. Classified foreground stars are plotted as lavender $\times$-symbols. 
The blue curve shows the locus of unreddened dwarf stars, the orange curve shows the locus of unreddened giants, and the dashed red curve shows the classical T Tauri locus \citep{CTTSNIR}. ({\it b}) $J$ versus $J-H$ color--magnitude diagram for the sources in panel ({\it a}). The blue curve shows the ZAMS with spectral types marked (at $d = 2.0$ kpc), and the brown dashed curves are 1 and 3 Myr isochrones from \citet{SDF00}. 
}
\label{fig:JHK}
\end{figure}

We automatically classified as PCM 152 X-ray sources with MIR counterparts identified as YSOs from SED modeling (SED\_flg = 0, Table~\ref{tab:xsed}).
In addition, among 137 X-ray sources with a NIR counterpart but either no MIR counterpart or no successful SED fits, 38 exhibited significant $K_S$-excess colors beyond their photometric errors (dots plotted to the right of the locus of reddened stars in Figure~\ref{fig:JHK}a) and were classified PCM. (All NIR counterparts exhibiting significant $K_S$ excess, irrespective of SED fits, are flagged in column 9 of Table~\ref{tab:xsed}.) Only a {\em manual} override to the classification (see~\ref{sec:override} below) can remove an X-ray detected YSO from the PCM sample.

\subsubsection{X-ray variability}\label{sec:variability}
Our continuous, nearly 100 ks ACIS integration provides a very good temporal baseline for detecting the rapid, high-amplitude X-ray flares frequently produced by young stars \citep{W05}. X-ray variability observed during the ${\sim}1$-day timeframe of our data hence provides very strong evidence favoring membership. 
Following \citet{CCCPClass,MYStIXPCM}, we use the $p_{\rm NV}$-value for the no-variability hypothesis, estimated using the Kolmogorov--Smirnov (K-S) statistic (defined as ProbKS\_single in our X-ray source catalog).
We flagged 75 sources with $p_{\rm NV}<0.005$ as ``definitely X-ray variable'' (column 16 of Table~\ref{tab:xsed}), and classified them as PCM, including 54 new PCMs not previously classified on the basis of IR excess (nine of which had no NIR or MIR counterpart at all). As in the case of X-ray detected YSOs above, only a manual override (Section \ref{sec:override}) can change the PCM classification of an X-ray variable source.

\begin{figure}[thp]
\epsscale{0.6}
\plotone{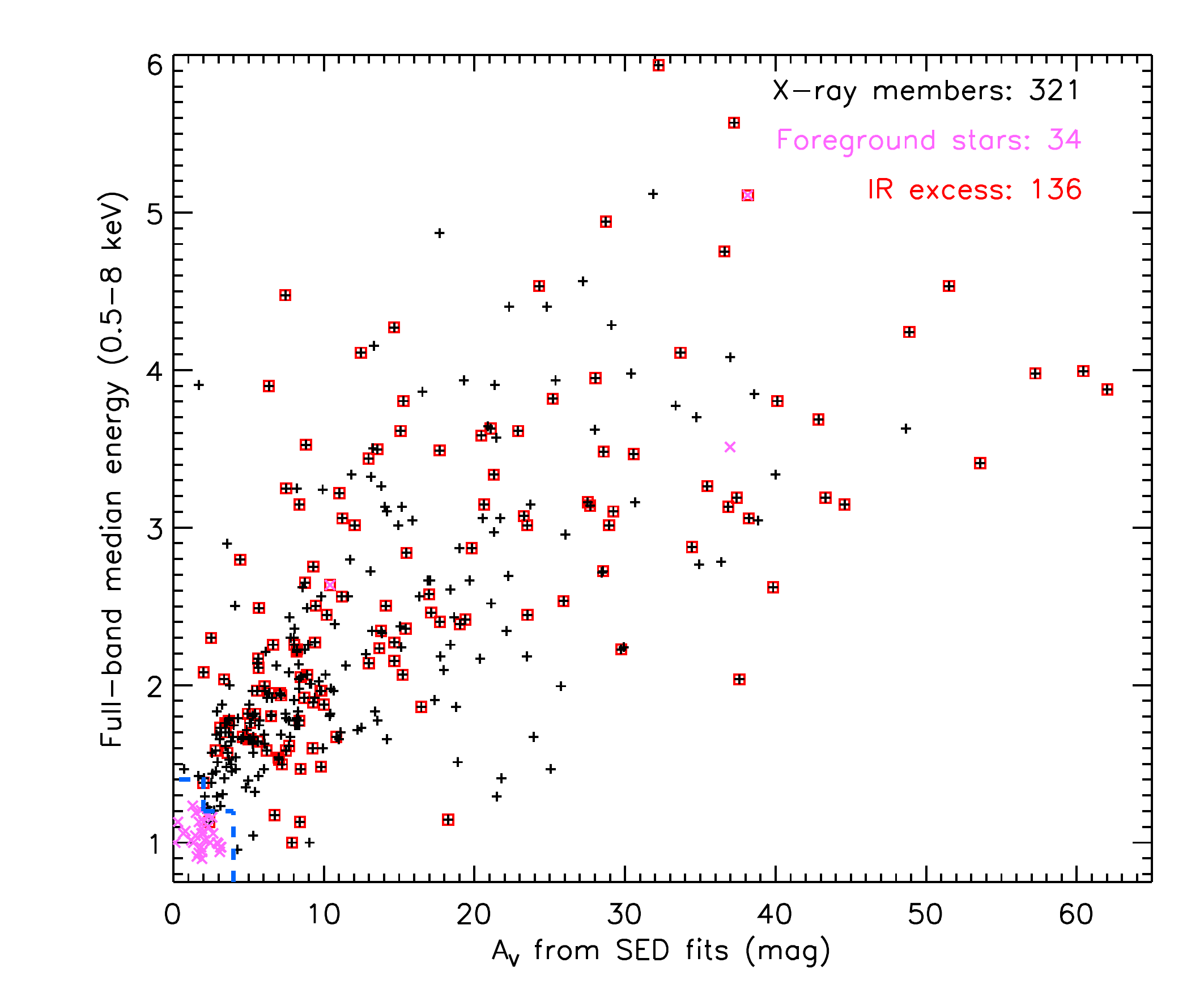}
\caption{Plot of full-band X-ray median energy ($E_{\rm med}$) versus visual extinction (\avsed) for 356 X-ray sources with ${>}4$ net counts and successful SED fitting of the IR counterparts (including 136 with IR excess; red boxes). The region to the lower left of the blue dashed lines is used to separate likely foreground stars (lavender $\times$-symbols) from X-ray PCMs (black crosses); see also
Fig.~\ref{fig:JHK}. 
Three sources ($\times$-symbols with high $E_{\rm med}$ and \avsed, two of which have IR excess) were manually classified as belonging to an obscured foreground cluster in one corner of the ACIS FOV (see Section~\ref{sec:override}).
}
\label{fig:MEvAv}
\end{figure}

\subsubsection{X-ray median energy and IR extinction}
Based on our expectation that the IRDC will screen out background contaminants, in particular older, field stars that are not X-ray bright to begin with, X-ray sources with IR counterparts consistent with reddened hot stars or pre-MS stars should be PCM, even if lacking IR excess.
Conversely, non-IR excess stars with low X-ray median energy ($E_{\rm med}$, reported in column 15 of Table~\ref{tab:xsed} for 632 X-ray sources with ${>}4$ net counts), low interstellar reddening to the IR counterpart, or both, are likely foreground. For IR counterparts with non- or marginal-IR excess SED classifications (SED\_flg = -2 or -1 in Table~\ref{tab:xsed}) foreground stars satisfy the following constraints (illustrated in Figure \ref{fig:MEvAv}, which plots $E_{\rm med}$ versus \avsed\ determined from our SED modeling):
\begin{displaymath}
\left(E_{\rm med} \le 1.2 ~{\rm keV~~AND~~} A_V^{\rm SED} \le 4~{\rm mag}\right)
~~{\rm OR}~~
\left(E_{\rm med} \le 1.4 ~{\rm keV~~AND~~} A_V^{\rm SED} \le 2~{\rm mag}\right).
\end{displaymath}
Additionally, we classify stars as foreground if they have
$A_V^{\rm SED}\le 4~{\rm mag}$
but ${\le}4$ net counts and hence no reliable $E_{\rm med}$ measurement. 
We hence identified 41 foreground stars that satisfied the above low-absorption/extinction criteria and 178 new PCMs that did not. 

The numerical cutoffs for $E_{\rm med}$ and $A_V$ used to separate foreground stars from PCM were determined iteratively, starting with expectations for the foreground reddening to M17 based on its distance and location in the Galactic plane \citep{GalacticAV} and typical $E_{\rm med}$ for unobscured, pre-MS stars \citep{CCCPClass,CCCPSim}.
We refined our selection criteria to visually isolate the loci of foreground stars versus PCMs in the $E_{\rm med}$ vs.\ \avsed\ plot of Figure~\ref{fig:MEvAv} and in the NIR color-color diagram and color-magnitude diagrams (CMD) shown in Figure~\ref{fig:JHK}.
The observed distributions of the 244 PCMs previously classified using IR excess or X-ray variability were important guides for visually determining the boundaries of the foreground loci, as these high-confidence PCMs are representative of the X-ray and IR properties of the larger M17 SWex young stellar population. Our ability to separate foreground stars from PCMs is not perfect, as evidenced by the ${\sim}10$ NIR counterparts to X-ray PCMs appearing blueward of the 3~Myr isochrone in Figure~\ref{fig:JHK}b, occupying a similar color space to foreground stars. As these sources represent only ${\sim}3\%$ of our X-ray PCMs, even if they are misclassified foreground stars they contribute a negligible level of contamination to our sample.

For X-ray sources without available \avsed, if a NIR counterpart is detected in all three bands we can compute the equivalent visual extinction $A_V^{\rm NIR}$ required to deredden the NIR to the nearest point on the unobscured MS locus (Figure~\ref{fig:JHK}a). This reddening measurement, commonly used in NIR studies, is generally appropriate for cool, pre-MS stars, but for hot stars with $A_V \ga 5$~mag $A_V^{\rm NIR}$ will underestimate the true extinction by ${\sim}5$~mag. Applying our cuts using $A_V^{\rm NIR}$ to 83 X-ray sources with NIR counterparts but no SED fitting identified 26 additional foreground stars and 59 additional PCMs.

Four X-ray sources had NIR counterparts, detected in all three bands, consistent with red giant colors (these sources are omitted from Figure~\ref{fig:JHK}a, but would fall above the locus of reddened giants). Evolved stars are generally not expected to be bright enough in X-rays to be detected by our observations. We left these sources unclassified (U), because they could be foreground, background, or even mismatches between X-ray and NIR sources. All four of these sources were located far off-axis, where the \chandra\ PSF is large, increasing the chances of a mismatch, and away from the IRDC lanes, where the extinction is lower and hence the density of background field giants in the IR images is higher.

Four X-ray sources with no IR counterpart of any kind but low $E_{\rm med}<1.2$~keV and no evidence for variability were classified as foreground stars.

\subsubsection{Other IR color criteria}
IR counterparts to X-ray sources with photometry in a combination of $N_{\rm data} < 4$ bands {\em other} than $JHK_S$ have no SED fitting results available and cannot be plotted on the color-color and color magnitude diagrams of Figure~\ref{fig:JHK}.
For such sources we cannot reliably distinguish between red colors caused by interstellar versus circumstellar reddening, but above a certain color threshold the distinction does not matter, as very red colors are evidence for PCM. We identified these thresholds by comparing the distributions of $H-K_S$ and $J-K_S$ colors for the remaining unclassified sources to those of sources previously classified as foreground versus PCM. We classified 23 new PCMs with $H-K_S\ge0.3$~mag (without $J$ detections; not counting one PCMs previously classified by X-ray variability) and 7 new PCMs classified with $J-K_S\ge 1$~mag (without $H$ detections). No previously unclassified sources had colors consistent with foreground stars using these color cuts.

We also identified 16 previously unclassified PCMs with MIR counterparts but no NIR counterparts (excluding one previously classified by X-ray variability). These sources were classified as PCM because a detection in the GLIMPSE Archive but not in the deep UKIDSS observations implies a highly-embedded X-ray source.

\subsubsection{Proximity to another PCM}\label{sec:PCMprox}
After all of the other classification steps were performed, we visually identified a number of unclassified X-ray sources located in several tight clusters that contained significant numbers of PCMs, coinciding with several compact MIR nebulae in M17 SWex (see Figure~\ref{fig:Xcat}). Two of these clusters are associated with ``Hub-N'' and ``Hub-S,'' dense nodes where multiple filaments intersect identified by B13. The clusters are too dense or MIR-bright to have point sources resolved by IRAC and too dense or embedded for all of the X-ray sources to have counterparts detected by UKIDSS.

To identify X-ray MCPMs within these small clusters or otherwise in close proximity to other PCMs, we computed the mean separation $s_4$ of each remaining unclassified source to its four nearest PCM neighbors. We found that the distribution of $s_4$ among the 241 remaining unclassified sources was highly skewed toward larger values of $s_4$. We thus rejected all $s_4$ values that were ${>}1$-sigma outliers above the mean, and recomputed the mean and variance ($\bar{s}_4=23.5\arcsec$ and $\sigma_{s_4}=8.8\arcsec$) from this sigma-clipped distribution. We note that these angular separations are sufficiently small to be generally insensitive to the large-scale variation in \chandra\ sensitivity with off-axis angle $\theta$, however this spatial filtering is biased against unclassified sources with large $\theta$ values, where the overall density of detected X-ray sources is relatively low. We set a somewhat arbitrary threshold of $s_4\le \bar{s}_4 - \sigma_{s_4}$ to assign the most highly-clustered unclassified sources to the PCM class; this criterion identified 41 new PCM, or 17\% of the remaining unclassified sources, in the closest proximity to PCMs classified using other criteria.


\begin{deluxetable}{llrrrl}
\tabletypesize{\scriptsize}
\tablecaption{Summary of X-ray Source Classification\label{Xclass}}
\tablewidth{0pt}
\tablehead{
  \colhead{Sec.} & \colhead{Criterion} & \colhead{PCM} & \colhead{Foreground} & \colhead{Unclassified} & \colhead{Description}
}
\startdata
A.3.1 & MIR Excess Emission    &  152 &   & 688 & {\em Spitzer} counterpart is candidate YSO \\
A.3.1 & $K_s$-Excess Colors    &   38 &   & 650 &  No {\em Spitzer} detection; $K_S$ excess in NIR counterpart \\
A.3.2 & X-ray Variability      &   54\tablenotemark{a} &   & 596 & ``Definitely Variable'' X-ray sources \\
A.3.3 & $E_{\rm med}$ and $A_V^{\rm SED}$ &   178 & 41 & 377 & X-ray median energy and extinction from SED fitting  \\ 
A.3.3 & $E_{\rm med}$ and $A_V^{\rm NIR}$ &   59  & 24 & 294 & X-ray median energy and extinction from $JHK_s$ colors \\ 
A.3.3 & $E_{\rm med}$ only      &       & 4  & 290  & Low X-ray median energy and no IR counterpart \\
A.3.4 & $H-K_S \ge 0.3$        & 23    &    & 267 & NIR color consistent with highly-reddened star \\
A.3.4 & $J-K_S \ge 1$          &  7    &    & 260 & NIR color consistent with highly-reddened star \\
A.3.4 & MIR only               &  16   &    & 244 & MIR counterpart but no NIR detection \\
A.3.5 & Spatial Proximity      &  41   &    & 203 & Clustered near previously-classified PCMs \\
A.3.6 & Manual override\tablenotemark{b} &  $-4$  & 3  &  203   & Associated with known foreground/background objects \\ 
\hline
& Totals                       & 564 & 72 & 203  & Final X-ray source classification tallies \\
\enddata
\tablecomments{The PCM and Foreground columns list number of sources newly classified by application of each successive criterion, while the Unclassified column shows the sources remaining unclassified at each step.}
\tablenotetext{a}{A total of 75 X-ray variable sources were identified, 21 of which were already classified as PCMs based on IR excess emission.}
\tablenotetext{b}{A single source, CXO J181837.88--170248.1, was classified as {\em background} by manual override and does not appear in the final classification tallies in this Table.}
\end{deluxetable}

\subsubsection{Manual override of individual source classifications}\label{sec:override}
Once the semi-automated classification procedure described above was complete, we checked for associations of X-ray sources with known astrophysical objects that might necessitate a manual change in classification. 

The three sources in the probable foreground embedded cluster G14.33--0.64, occupying the cut-out northeast corner of the ACIS field in Figure \ref{fig:ISED}, were re-classified from PCM to foreground. Two are candidate X-ray detected YSOs, one of which is additionally a remarkable, highly-absorbed X-ray source (see Section~\ref{sec:surprising} for details).

The brightest source in our sample, by far, is CXO J181837.88--170248.1, with 3054 net total-band counts and definite variability ($p_{\rm NV}=0$). This source is SAX J1818.6--1703 \citep{SAXdisc}, a high-mass X-ray binary with an OB supergiant companion in a 30-day orbit \citep{ZHC09,B09SAX}.
This source was previously observed for 9.8 ks using the ACIS-S array \citep{SAXChandra}, and while its distance is not known, its strongly-absorbed, power-law spectrum suggests that it is a background object observed through the outer envelope of M17 SWex.

The net effect of our manual classification overrides was to decrease the PCM class by 4 sources, classify one additional foreground star, and assign a single source to the background class. No new PCM classifications were made using manual override.

\section{Details of SED Modeling for X-ray Detected, Non-IR excess Sources}\label{append:construct}

In this Appendix we describe the the stellar models and details of the methodology used to analyze the results of fitting SED models to the diskless, X-ray-selected young stellar population.

\subsection{A New Grid of $10^5$ SED Models for ``Naked'' T Tauri Stars, IMPS and Main-Sequence Stars}

We have produced a set of stellar SED models appropriate for stars that show no evidence of IR excess due to circumstellar disks, including ``naked'' T Tauri Stars, IMPS, and main-sequence stars. These models include only stellar parameters and none of the various parameters related to circumstellar material used in the RW06 YSO models. The six parameters of the naked stellar grid are listed in Table~\ref{tab:pars}. Unlike the YSO models, the naked stellar models are spherically symmetric, so inclination angle is not a parameter.\footnote{Unlike the RW06 and similar YSO model SEDs, for the naked stellar models there is no need to perform radiation transfer calculations to propagate photons through circumstellar dust. Once a set of model stellar spectra and pre-MS evolutionary tracks are adopted, the naked stellar models are not computationally expensive.}

\begin{deluxetable}{llccc}
\tabletypesize{\small}
\tablewidth{0 pt}
\tablecaption{ \label{tab:pars}
 Parameters in the Naked Stellar SED Models}
\tablehead{Parameter & Description & Unit & Minimum & Maximum}
\startdata
$t_{\star}$  & Evolutionary age & yr & $10^4$ & $10^{10}$ \\
$M_{\star}$ & Mass & \Msun & 0.1 & 60 \\
$L_{\rm bol}$ & Bolometric luminosity & \Lsun & $8.2\times 10^{-4}$ & $5.4\times 10^{5}$ \\
$T_{\rm eff}$ & Effective temperature & K & 2600 & $4.8\times 10^4$ \\ 
$R_{\rm eff}$ & Effective radius & $R_{\sun}$ & 0.11 & 67 \\
$\log_{10}{g}$ & Surface gravity & m s$^{-2}$ &  1.9 & 5.3
\enddata
\end{deluxetable}

Stellar masses $M_{\star}$ were uniformly sampled in logarithmic space between the minimum ($M_{\rm min}=0.1$~\Msun) and maximum ($M_{\rm max}=60$~\Msun) values (Table~\ref{tab:pars}). 
Evolutionary ages $t_{\star}$ were similarly sampled between $t_{\rm min}=10^4$~yr and $t_{\rm max}=10^{10}$~yr. If the resulting value for $t_{\star}$ exceeded the combined pre-MS and MS lifetime for a star of mass $M_{\star}$, the age was resampled until it reached an acceptable value. 
(Fig.~\ref{fig:params}b), which terminates at the MS turnoff age, since we do not include models for post-MS stars. 
We note that the RW06 YSO models used a different probability distribution (their Equation 2) for sampling $t_{\star}$, producing a slight increase in model density toward older ages.

Following RW06, the stellar effective temperature $T_{\rm eff}$ and radius $R_{\rm eff}$, and hence bolometric luminosity $L_{\rm bol}$, were derived for each model combination of $M_{\star}$ and $t_{\star}$ by interpolating pre-MS evolutionary tracks, specifically \citet{BM96} for $M_{\star} \ge 9$~\Msun, \citet{SDF00} for $M_{\star} \ge 9$~\Msun, and a combination of the two for 7~\Msun$< M_{\star} <9$~\Msun. These particular pre-MS evolutionary models are therefore a built-in {\em assumption} of our naked stellar models, used to transform the physical parameters most closely tied to observables, $T_{\rm eff}$ and $L_{\rm bol}$, into the fundamental parameters of $M_{\star}$ and $t_{\star}$. A different set of pre-MS evolutionary models \citep[those of][for example]{Dotter} could be employed to provide a different transformation. We have not yet explored the effects of adopting different evolutionary models, however the models we have adopted remain widely used in both Galactic and extragalactic studies.

The final parameter in Table~\ref{tab:pars}, $\log_{10}{g}$, is readily derived from $M_{\star}$ and $R_{\rm eff}$. To construct the model SEDs, each model combination of $T_{\rm eff}$ and $\log_{10}{g}$ is used to select the appropriate spectrum from the set of \citet{Kurucz} model atmospheres, and then $L_{\rm bol}$ provides the normalization for the SED. The naked stellar model SEDs are convolved with the appropriate broadband filters and fit to the data using the \citet{fitter} fitting tool, as described in \ref{IRE} above.

\subsection{Weighting SED Model Fits Using Age-Based Priors}


Following the approach used in our previous work (e.g.\ PW10, \citealp{CCCPYSOs}, and P13), to constrain the parameters returned by fitting either the RW06 YSO or our new naked stellar SED models to photometry data, we define the set $i$ of well-fit models using 
\begin{equation}
  \frac{\chi^2_i - \chi^2_0}{N_{\rm data}} \le 2,
\end{equation}
where $\chi^2_0/N_{\rm data}$ is the goodness-of-fit parameter for the best-fit model, normalized to the number of photometry datapoints used in the fit (column 9 of Table~\ref{tab:ised} and column 12 of Table~\ref{tab:xsed}).
We then $\chi^2$-weight each model fit as
\begin{equation}
  W_i(\chi^2) = e^{-\chi^2_i/2}.
\end{equation}
For MIRES fit with YSO models, the probability of a given model fit is simply $P_i = P_nW_i(\chi^2)$, with normalization constant $P_n$ (P13).

For the naked stellar models fit to non-MIRES counterparts to X-ray sources, the fitting procedure yields relatively poor constraints on $T_{\rm eff}$, which is degenerate with the external parameter $A_V$. Ideally, one would obtain spectra of the stellar populations to derive $T_{\rm eff}$ independently, and then SED fitting would provide strong constraints on extinction and luminosity, however at present it is prohibitively expensive to perform NIR spectrocopy on hundreds of obhjects in distant, heavily-obscured regions such as M17 SWex. We therefore leverage two pieces of {\em external} knowledge, namely the range of evolutionary ages consistent with a source being both {\em diskless} and {\em detectible in X-rays} to weight the probability distributions according to the $t_{\star}$ model parameter. This is conceptually similar to adopting a Bayesian prior, although we are not performing Bayesian statistical analysis. The two new weighting factors, described below, are $W_i(\tau_d)$ and $W_i(\tau_X)$. The probability of an individual naked stellar model fit is thus
\begin{equation}
  P_i = P_n W_i(\chi^2) W_i(\tau_d) W_i(\tau_X),
\end{equation}
where $P_n$ is again chosen as a normalization constant such that $\sum{P_i}=1$.

\subsubsection{Lack of an Inner Disk Disfavors Very Young Ages}

We use disk lifetimes to constrain ages of diskless PMS stars to the extent that we assume inner disks that would produce detectible MIR excess emission become disfavored for $t_{\star}<\tau_d(m)$, where the disk lifetime is defined using the PW10 power-law parameterization as
\begin{equation}
  \tau_d(M_{\star}) =
  \begin{cases}
    \tau_d(M_c)\left(\frac{M_{\star}}{M_c}\right)^{-\delta},& M_c< M_{\star} < M_c' \\
    \tau_d(M_c),& M_{\star} \le M_c
  \end{cases}
\end{equation}
(see Section~\ref{sec:disks}). 
We assume $M_c = 1$~\Msun\ with $\tau_d(M_c)=3$~Myr, the ``canonical'' disk lifetime for low-mass YSOs \citep{HLL01} and $M_c'=10$~\Msun\ with $\tau_d(M_c')=0.1$~Myr for massive YSOs \citep{RMS11b}. These constraints together yield the power-law index $\delta=1.5$ for the mass-dependent disk lifetime. We also adopt an uncertainty on $\tau_d(m)$ of $\sigma_d(m)=0.2\tau_d(m)$ on the disk lifetimes, based on the allowable variation in $\delta$ that would still be consistent with our observed values of $\Gamma_{\rm YMF}$ (see Figure~\ref{fig:YMFs} and Section~\ref{sec:disks}).

We assume that the actual lifetimes among real YSOs are normally distributed relative to the mass-dependent average disk lifetime $\tau_d(m)$. We hence weight the model fits according to evolutionary age $t_{\star,i}$ using the Gaussian cumulative distribution function computed via the error function,
\begin{equation}\label{eq:diskweight}
  W_i(\tau_d) = \frac{1}{2}\left[1+{\rm erf}\left(\frac{t_{\star,i}-\tau_d/2}{\sigma_d\sqrt{2}}\right)  \right]. 
\end{equation}
Recall that for a given model $i$, $\tau_d$ and hence $\sigma_d$ are functions of $M_{\star,i}$ (Equation~\ref{eq:diskfrac}). Checking the limiting behaviors of this weighting function, we find $W_i(\tau_d)=0$ for $t_{\star,i}=0$ (all stars are born with disks) and $W_i(\tau_d)= 1$ for $t_{\star,i}\ge \tau_d(M_{\star,i})$ (at evolutionary ages beyond the disk lifetime, we expect to find plenty of stars without disks).

For massive models ($M_{\star,i}>8$~\Msun) we set $W_i(\tau_d)=1$ (no weighting) regardless of $t_{\star,i}$, but in practice this step proved unnecessary, because effectively zero model fits to luminous, non-MIRES SEDs returned sufficiently young ages ($t_{\star,i} \la 0.1$~Myr) to produce $W_i(\tau_d) < 1$ according to Equation~\ref{eq:diskweight}.

\subsubsection{X-ray Detection Disfavors Older Ages}
Once a star arrives on the ZAMS, it is impossible to place a useful upper bound on its age using broadband photometry alone. For all but the most massive stars, the main-sequence turnoff age (the maximum $t_{\star}$ allowed by our models) is many orders of magnitude older than the stellar population in a star-forming region such as M17 SWex. This presents a problem, because if {\em any} main-sequence model provides an acceptable fit to a given source SED, then {\em all} main-sequence models at that mass will provide acceptable fits, biasing the distribution of $t_{\star,i}$ toward unrealistically old ages.

The fact that our sources have X-ray detections mitigates this SED model bias. \citet{PF05} found that the X-ray luminosity of pre-MS stars  in the mass range 0.5~\Msun$< m <1.2$~\Msun\ decays rapidly with age, as $L_{t,c}\propto t_{\star}^{-0.75}$. This explains empirically why X-ray observations are so effective at detecting young stellar populations against overwhelming contamination from older field stars in IR images and also provides a useful, quantitative weighting constraint for interpreting our SED model fits. As with most relations between X-ray luminosity and other stellar properties, there is a very large scatter about the trend, of roughly an order of magnitude in $L_{t,c}$. The relative probability, as a function of $t_{\star}$, that we can detect in X-rays a star of a particular mass at a given evolutionary age can hence be approximated as the {\em fraction} of stars with mass $M_{\star}$ and age $t_{\star}$ that happen to have $L_{t,c}$ above the sensitivity limit of the X-ray observation. This fraction decreases approximately as $L_{t,c}$, and we hence model the time-decay in probability of an X-ray detection as
\begin{equation}\label{eq:xweight}
W_i(\tau_X) = 
\begin{cases}
  \left(\frac{t_{\star,i}}{\tau_X}\right)^{-0.75}, & t_{\star,i} > \tau_X \\
  1, & t_{\star,i} \le \tau_X
\end{cases}
\end{equation}
where the constant $\tau_X=1$~Myr is the evolutionary age beyond which the steep decay in luminosity reported by \citet{PF05} becomes clearly observable (their Fig.~1).

It is strictly appropriate to apply $W_i(\tau_X)$  only for T Tauri stars in the mass range 0.5~\Msun$< m <1.2$~\Msun.  Our IR observations are insufficiently deep to detect most such low-mass stars in M17 SWex, as evidenced by Figure \ref{fig:YMFs}. But the intermediate-mass, A- and B-type ZAMS stars in our sample are expected to be X-ray quiet, so we may assume that unresolved, T Tauri binary companions to the IR-detected primary are responsible for the detected X-ray source \citep{CCCPcomps}. In such cases, $W_i(\tau_X)$ is appropriate for modeling intermediate-mass stars, 
since young, secondary companions detected in X-rays provide age constraints on their coeval, intermediate-mass primaries.

Because the X-ray emission processes for massive, OB stars are fundamentally different from low-mass, pre-MS stars, we set $W_i(\tau_X)=1$ (no weighting) for all models with $M_{\star,i}>8$~\Msun. We do not base any of our age constraints on the high-mass, non-MIRES population.

\end{document}